\documentclass[preprintnumbers, prd, floatfix,twocolumn,preprintnumbers, letterpaper, superscriptaddress,nofootinbib]{revtex4-1}

\usepackage{amsfonts}
\usepackage{amsmath}
\usepackage{amssymb,epsf}
\usepackage{latexsym}
\usepackage{graphicx,epsfig}
\usepackage{amssymb}
\usepackage{float}
\usepackage{subfigure}
\usepackage{epstopdf}
\usepackage[colorlinks=true,citecolor=blue,linkcolor=blue,urlcolor=black]{hyperref}
\usepackage{dcolumn}
\usepackage{psfrag}
\usepackage{wrapfig}
\usepackage{makeidx}
\usepackage{epsf}
\usepackage{color}
\usepackage{multirow}
\usepackage{tikz}

\graphicspath{{Images/}}

\definecolor{lime}{HTML}{A6CE39}
\newcommand{\orcidicon}{%
    \begin{tikzpicture}
    \draw[lime, fill=lime] (0,0)
        circle [radius=0.16]
        node[white] {{\fontfamily{qag}\selectfont \tiny ID}};
    \draw[white, fill=white] (-0.0625,0.095)
        circle [radius=0.007];
    \end{tikzpicture}   \hspace{-2mm}
}
\newcommand\orcidFrancisco{{\href{https://orcid.org/0000-0002-9388-8373}{\orcidicon}}}
\newcommand\orcidMahdi{{\href{https://orcid.org/0000-0003-1196-9493}{\orcidicon}}}

\begin{document}

\title{Shadow, deflection angle and quasinormal modes of Born-Infeld charged black holes}
\author{Khadije Jafarzade}
\email{khadije.jafarzade@gmail.com}
\affiliation{Department of Physics, Faculty of Science, University of Mazandaran, Babolsar 47416-95447, Iran}
\author{Mahdi Kord Zangeneh\orcidMahdi\!\!}
\email{mkzangeneh@scu.ac.ir}
\affiliation{Physics Department, Faculty of Science, Shahid Chamran University of Ahvaz,
Ahvaz 61357-43135, Iran}
\author{Francisco S. N. Lobo\orcidFrancisco\!\!}
\email{fslobo@fc.ul.pt}
\affiliation{Instituto de Astrof\'isica e Ci\^encias do Espa\c{c}o, Faculdade de
Ci\^encias da Universidade de Lisboa,\\ 
Departamento de F\'isica, Faculdade de
Ci\^encias da Universidade de Lisboa,\\ 
Edif\'icio C8, Campo Grande,
P-1749-016, Lisbon, Portugal}
\date{\today}

\begin{abstract}
In this paper, we consider black holes in the consistent Aoki-Gorji-Mukohyama theory
of the four-dimensional Einstein-Gauss-Bonnet (4D EGB) gravity in the presence of Born-Infeld (BI) nonlinear electrodynamics. We study several optical features of these black holes such as the shadow radius, energy emission rate and deflection angle, and analyse the effect of the coupling constants, the electric charge and cosmological constant on the considered optical quantities. Furthermore, we also employ the connection between the shadow radius and quasinormal modes (QNMs) and investigate small scalar perturbations around the black hole solution. We show that the variation of the parameters of the theory provide specific signatures on the optical features of the BI charged black hole solution, thus leading to the possibility of directly testing this consistent Aoki-Gorji-Mukohyama 4D EGB black hole model by using astrophysical observations.
\end{abstract}

\maketitle

\section{Introduction}

Recently, a novel four-dimensional Einstein-Gauss-Bonnet (4D EGB) gravity
was proposed by Glavan and Lin \cite{EGB0} by rescaling the Gauss-Bonnet
(GB) coupling $\alpha \rightarrow \tilde{\alpha}/(D-4)$, and then defining
the four-dimensional theory as the limit $D\rightarrow 4$, at the level of
the field equations. In this manner, the GB term yields a non-trivial
contribution to the gravitational dynamics. It has also been discussed
whether this regularized theory is reliable \cite%
{EGB17a,EGB17b,EGB17c,EGB17d,EGB18,Arrechea:2020evj,Gurses:2020rxb,Arrechea:2020gjw}
and several remedies have been suggested to overcome the objections raised 
\cite{EGB18,EGB20,EGB21,EGB22}. It is worthwhile to mention that these
theories still hold the spherically symmetric 4D black holes as valid
solutions \cite{EGB20,EGB21}. In \cite{Aoki:2020lig}, the consistent Aoki-Gorji-Mukohyama theory
of 4D EGB gravity was proposed and several applications were further studied 
in \cite{Aoki:2020iwm,Aoki:2020ila}. In fact, the novel 4D
EGB theory of gravity has recently attracted much attention in a wide
context of black hole physics \cite%
{Fernandes:2020rpa,Konoplya:2020qqh,Kumar:2020uyz,Kumar:2020xvu,Zhang:2020qew,Shaymatov:2020yte}%
, and applications, such as, in black hole shadows \cite%
{Konoplya:2020bxa,Guo:2020zmf,EslamPanah:2020hoj,Zeng:2020dco},
thermodynamics issues \cite{Hegde:2020xlv,Singh:2020xju}, rotating solutions 
\cite{Kumar:2020owy,Wei:2020ght,Konoplya:2020fbx}, stability issues \cite%
{Ghosh:2020vpc,Konoplya:2020juj,HosseiniMansoori:2020yfj,Wei:2020poh,Konoplya:2020cbv,Zhang:2020sjh,Aragon:2020qdc,Liu:2020evp}%
, gravitational lensing \cite{Islam:2020xmy}, thin accretion disk around 4D
EGB black holes \cite{Liu:2020vkh}, causal structure \cite%
{Ge:2020tid,Dadhich:2020ukj}, and black hole solutions coupled to nonlinear
electrodynamics (NLED) \cite%
{DVSingh,Ghosh:2018bxg,DVSingh,Jusufi:2020qyw,Jafarzade:2020ilt,Yang:2020jno}.

Motivated by recent astrophysical observations, namely, the first image of a
supermassive black hole candidate in the center of M87 galaxy captured by
the Event Horizon Telescope project \cite{KAkiyama1,KAkiyama2}, we are
interested in performing an in-depth analysis of the optical features of
black hole solutions. More specifically, we consider the solution obtained
in 4D EGB gravity minimally coupled to Born-Infeld (BI) electrodynamics in
the presence of a negative cosmological constant, provided in \cite%
{Yang:2020jno}. In fact, BI electrodynamics is a well-known NLED theory, proposed by Born and Infeld in 1934 \cite{BornInfeld}%
, in order to regularize the ultraviolet divergent self-energy of a
point-like charge in classical dynamics.
Moreover, it was shown later that
this nonlinear theory may come from the low energy limit of open superstring
theory \cite{fradkin1985effective,metsaev1987born,bergshoeff1987born}. In
addition, experiments of photon-photon interaction have proposed a nonlinear
model for electrodynamics in vacuum \cite{nonlin1,nonlin2,nonlin3,nonlin4,nonlin5}.
The effects of NLED from a cosmological viewpoint have been studied, for instance, in \cite{Vargas1a,Vollick1a}. It has also been shown that NLED theories can remove both the big bang and black hole singularities \cite{Corda1b,DeLorenci,Cuesta1a}.
The shadow of nonregular black holes in the presence of the Euler-Heisenberg NLED, which is effectively the low-energy limit of BI electrodynamics, has been determined in \cite{EHTM8}, and comparing the results with the ones of the M87 imaged by the Event Horizon Telescope Project, the first astrophysical constraints on nonlinear electrodynamics has been presented. Hence, it is a significant expectation that NLED may affect astrophysical characteristics such as the shadow radius of a charged black hole.
In the context of 4D EGB gravity,
several generalized solutions were obtained in the literature such as
generalized Reissner-Nordstrom black hole with a Maxwell electric field
coupling to the 4D EGB gravity \cite{Fernandes:2020rpa}, Bardeen-like \cite%
{Kumar:2020uyz} and Hayward-like \cite{Kumar:2020xvu} black holes by
interacting the new gravitational theory with nonlinear electrodynamics, and
AdS black holes in the novel 4D EGB gravity coupled to exponential nonlinear
electrodynamics \cite{DVSingh,Ghosh:2018bxg,Jafarzade:2020ilt}. Thus, in
this work, we specifically analyse the shadow geometrical shape, the energy
emission rate, the deflection angle and quasinormal modes in the BI charged
black hole solution obtained in \cite{Yang:2020jno}. The behavior of photon
orbits around black holes in the presence of a variety of nonlinear
electrodynamics have also been studied in \cite{Xu:2019yub,Habibina:2020msd}.
In addition to various theories of black holes,
the effects of NLED theories would become quite important in super-strongly magnetized compact objects, such as pulsars, particular neutron stars, magnetars and strange quark magnetars  \cite{Mosquera1a,Mosquera1b,Bialynicka1b}.

This paper is organised in the following manner: In section \ref{sec:II}, we
present the consistent Aoki-Gorji-Mukohyama action and the specific black hole solution of 4D EGB gravity minimally coupled to BI electrodynamics in the presence of a negative cosmological constant. In section \ref{sec:III}, we
consider the motion of a free photon in the black hole background, and write
out the specific condition for a circular null geodesic. Then, we analyse
the photon sphere radius by varying the black hole parameters in subsection.
We also consider the radius of the shadow and examine the effect of the model’s
parameters and analyse the energy emission rate. In
section \ref{sec:IV}, we study the connection between the radius of the
shadow and quasinormal modes and consider small perturbations around the
black holes. In section \ref{sec:V}, we consider the light deflection around
the black hole solution. Finally, in section \ref{sec:conclusion}, we
conclude by summarizing and discussing our results.


\section{Consistent Aoki-Gorji-Mukohyama theory of the 4D EGB theory and black hole solution}

\label{sec:II} 

The consistent Aoki-Gorji-Mukohyama theory of 4D EGB gravity has been proposed in \cite%
{Aoki:2020lig} and further perused in \cite{Aoki:2020iwm,Aoki:2020ila}.
Since this consistent version of the theory does not respect time
diffeomorphism, it is natural to use the $4$-dimensional Arnowitt-Deser-Misner
(ADM) metric of the form 
\begin{equation}
ds^{2}=-N^{2}dt^{2}+\gamma _{ij}(dx^{i}+N^{i}dt)(dx^{j}+N^{j}dt)\,,
\label{admmet}
\end{equation}%
where $N$, $N^{i}$ and $\gamma _{ij}$ are the lapse function, the shift
vector and the spatial metric, respectively. The well-defined gravitational
action in the presence of a negative cosmological constant is then given by 
\cite{Aoki:2020lig}%
\begin{equation}
S_{\mathrm{g}} =\frac{M_{\mathrm{Pl}}^{2}}{2} \int dtd^{3}xN\sqrt{\gamma }%
\left[-2\Lambda +2R-\mathcal{M}+\tilde{\alpha}R_{\mathrm{4DGB}}^{2}\right], 
\label{EGBact} 
\end{equation}
where
\begin{eqnarray}
R_{\mathrm{4DGB}}^{2} &\equiv & \frac{1}{2}\Big[8R^{2}-4R\mathcal{M}-\mathcal{%
M}^{2} \notag \\
&& -\frac{8}{3}\big(8R_{ij}R^{ij}-4R_{ij}\mathcal{M}^{ij}-\mathcal{M}_{ij}%
\mathcal{M}^{ij}\big)\Big],  \notag
\end{eqnarray}%
in which 
\begin{equation*}
\mathcal{M}_{ij}\equiv R_{ij}+\mathcal{K}^{k}{}_{k}\mathcal{K}_{ij}-\mathcal{%
K}_{ik}\mathcal{K}^{k}{}_{j},\hspace{1cm}\mathcal{M}\equiv \mathcal{M}%
^{i}{}_{i}\,,
\end{equation*}%
and 
\begin{equation*}
\mathcal{K}_{ij}\equiv \frac{1}{2N}(\dot{\gamma}_{ij}-2D_{(i}N_{j)}-\gamma
_{ij}D^{2}\lambda _{\mathrm{GF}})\,.
\end{equation*}%
Here, $M_{\mathrm{Pl}}=1/\sqrt{8\pi G}$ is the reduced Planck mass, $G$ is
the Newton gravitational constant and $\tilde{\alpha}$ is the rescaled
Gauss-Bonnet coupling constant. The factor $\lambda _{\mathrm{GF}}$ is the Lagrange multiplier related to a gauge-fixing constraint (we refer the reader to Ref. \cite{Aoki:2020lig} for more details).
In addition, $D_{i}$ is the covariant
derivative compatible with the spatial metric and dot denotes the derivative
with respect to $t$. We will also set $\Lambda $ to $-3/l^{2}$. One of the
properties of the consistent Aoki-Gorji-Mukohyama theory introduced by the action (\ref{EGBact}) is that if a $D$-dimensional solution of EGB gravity has vanishing Weyl tensor of the
spatial metric $\gamma _{ij}$ and the Weyl piece of $2K_{i[m\mid }K_{j\mid n]}$,
where $K_{ij}=\frac{1}{2N}(\dot{\gamma}_{ij}-2D_{(i}N_{j)})$, the $%
D\rightarrow 4$ limit of this solution is a solution of the well-defined theory of 4D EGB theory as well \cite{Aoki:2020lig,Aoki:2020iwm,Aoki:2020ila}. Furthermore, the
action (\ref{EGBact}) presents a consistent 4D EGB theory uniquely up to a
choice of the gauge-fixing constraint imposed by the Lagrange multiplier $%
\lambda _{\mathrm{GF}}$.

In this paper, our aim is to study optical features of Born-Infeld
charged black hole solutions of the consistent Aoki-Gorji-Mukohyama theory of 4D EGB gravity. Therefore, we minimally couple the gravitational Lagrangian to the BI electrodynamics Lagrangian given by
\begin{equation}
\mathcal{L}_{BI}=4\beta ^{2}\left( 1-\sqrt{1+\frac{2F}{\beta ^{2}}}\right) .
\label{BI_action}
\end{equation}
where $F=\frac{1}{4}F_{\mu\nu }F^{\mu \nu }$, with $F_{\mu \nu }=\partial _{\mu }A_{\nu }-\partial
_{\nu }A_{\mu }$ and $A_{\mu }$ is the potential. Here $\beta >0$
denotes the BI parameter which is the maximum of the electromagnetic field
strength.

Our desired solution namely BI charged black hole solution of 4D EGB
gravity, has already been obtained in \cite{Yang:2020jno}. However, we need
to show that this solution is a solution of the consistent Aoki-Gorji-Mukohyama theory (\ref{EGBact}).
As pointed out above, we just need to confirm that the Weyl tensor of
the spatial metric and Weyl piece of $2K_{i[m\mid }K_{j\mid n]}$ vanish for $%
D$-dimensional black hole solution of EGB gravity minimally coupled to BI
electrodynamics. This solution is a static and spherically symmetric
solution with just one metric function $f\left( r\right) $ \cite%
{Yang:2020jno,Aiello:2004rz,Zou:2010tv}. Hence, we consider the general form
of this kind of solution%
\begin{equation}
ds^{2}=-f(r)dt^{2}+\frac{dr^{2}}{f(r)}+r^{2}d\Omega _{D-2}^{2},
\label{metric}
\end{equation}%
where%
\begin{equation*}
d\mathbf{\Omega }_{D-2}^{2}=d\theta _{1}^{2}+\sum\limits_{i=2}^{D-2}d\theta
_{i}^{2}\prod\limits_{j=1}^{i-1}\sin ^{2}\left( \theta _{j}\right) ,
\end{equation*}%
is the line element of $(D-2)$-dimensional unit sphere. Comparing metric (%
\ref{metric}) with ADM metric (\ref{admmet}), one confirms that $N^{2}=f(r)$%
, $N^{i}=0$ and%
\begin{eqnarray}
&\gamma _{ij}=\mathrm{diag}\left( \frac{1}{f(r)},r^{2},r^{2}\sin ^{2}\left(
\theta _{1}\right) ,r^{2}\sin ^{2}\left( \theta _{1}\right) \sin ^{2}\left(
\theta _{2}\right) ,\cdots \right) . \nonumber \\  \label{gammij}
\end{eqnarray}%
Consequently, the condition $2K_{i[m\mid }K_{j\mid n]}=0$ where $K_{ij}=%
\frac{1}{2N}(\dot{\gamma}_{ij}-2D_{(i}N_{j)})$ is automatically satisfied.
Moreover, our calculations show that for a spatial metric $\gamma _{ij}$ of
the form given in Eq. (\ref{gammij}), the Weyl tensor also vanishes. Thus, the $%
D\rightarrow 4$ limit of any $D$-dimensional solution of EGB gravity of the
form given in Eq. (\ref{metric}), including the BI one, is a solution of the
consistent Aoki-Gorji-Mukohyama theory of 4D EGB gravity defined by (\ref{EGBact}) minimally coupled to
matter fields as well.

Varying the total action with respect to the gauge field, $A_{\mu }$,
provides the following equation of motion 
\begin{equation}
\partial _{\mu }\left( \frac{\sqrt{-g}F^{\mu \nu }}{\sqrt{1+2F/\beta ^{2}}}%
\right) =0\,.  \label{BIeom}
\end{equation}%
With ansatz (\ref{metric}), one could immediately integrate the
electrodynamics equation of motion (\ref{BIeom}) with the gauge field being $%
A=\Phi (r)dt$ and obtain
\begin{equation}
\Phi (r)=\frac{Q}{r}{~}_{2}F_{1}\left( \frac{1}{4},\frac{1}{2},\frac{5}{4},-%
\frac{Q^{2}}{\beta ^{2}r^{4}}\right) \,,  \label{solution_phi}
\end{equation}%
for $D=4$, in which $Q$ is the electric charge and $_{2}F_{1}$ is the
hypergeometric function. Then, the metric function $f\left( r\right) $ is
given by \cite{Yang:2020jno}%
\begin{eqnarray}
f(r) &=&1+\frac{r^{2}}{2\alpha }\Bigg\{1\pm \Bigg[1+4\alpha \Bigg(\frac{2M}{%
r^{3}}-\frac{1}{l^{2}}  \notag \\
&&-\frac{2\beta ^{2}}{3}\bigg(1-\sqrt{1+\frac{Q^{2}}{\beta ^{2}r^{4}}}\bigg)-%
\frac{4Q}{3r^{3}}\Phi (r)\Bigg)\Bigg]^{\frac{1}{2}}\Bigg\}.
\label{function-a}
\end{eqnarray}

Equation (\ref{function-a}) corresponds to two branches of solutions
depending on the choice of the sign. It can be shown that the positive-sign
branch reduces to a RN-AdS solution with a negative gravitational mass and
imaginary charge, and only the negative-sign branch recovers the proper
RN-AdS limit (we refer the reader to Ref. \cite{Yang:2020jno} for more
details). Thus, since only negative branch leads to a physically meaningful
solution, we will limit our discussions to this branch of the solution.

\section{Null geodesics and photon sphere}\label{sec:III}


In this section, we analyse the motion of a free photon in the black hole background (\ref{function-a}). Due to the  spherical symmetry of the spacetime, we consider that the motion is limited to the equatorial plane $\theta =\pi/2$, without a significant loss of generality. Thus, the Lagrangian for a photon can be expressed as
\begin{equation}
2\mathcal{L}=-f(r)\dot{t}^{2}+\frac{\dot{r}^{2}}{f(r)}+r^{2}\dot{\phi}^{2},
\label{EqLagrangian}
\end{equation}
where the overdot denotes a derivative with respect to the affine parameter. Taking into account Eq.
(\ref{EqLagrangian}), the equations of motion are given by
\begin{equation}
\dot{t}=-\frac{p_{t}}{f(r)},\qquad \dot{r}=p_{r}f(r), \qquad \dot{\phi}=\frac{p_{\phi}}{r^{2}},  \label{Eqmotion}
\end{equation}
where $ p $ is the generalized momentum defined by $ p_{\mu}=\partial \mathcal{L}/\partial \dot{x}^{\mu}=g_{\mu\nu}\dot{x}^{\nu} $. 

The Hamiltonian for this system reads
\begin{equation}
H=\frac{1}{2}g^{\mu\nu}p_{\mu}p_{\nu}=\frac{1}{2}\left[-\frac{p_{t}^{2}}{f(r)}+f(r)p_{r}^{2}+\frac{p_{\phi}^{2}}{r^{2}}\right]=0.
\label{EqHamiltonian}
\end{equation}
Since the Hamiltonian is independent of the coordinates $t$ and $\phi$, one can consider $ p_{t}\equiv -E $ and $ p_{\phi}\equiv L $ as constants of motion, where $ E $ and $ L $ are physically interpreted as the energy and angular momentum of the photon; note that $p_r$ is the radial momentum. Using the equations of motion and these two conserved quantities, one can rewrite the null geodesic equation as follows
\begin{equation}
\dot{r}^{2}+V_{\rm eff}(r)=0 , \label{Eqpotential}
\end{equation}
where the effective potential is 
\begin{equation}
 V_{\rm eff}(r)=f(r)\left[ \frac{L^{2}}{r^{2}}-\frac{E^{2}}{f(r)}\right] .  \label{Eqpotential12}
\end{equation}

For a circular null geodesic, the effective potential satisfies  the following conditions, simultaneously
\begin{equation}
V_{\rm eff}(r)\vert_{r=r_{p}}=0,\qquad \frac{\partial V_{\rm eff}(r)}{\partial r}\Big\vert_{r=r_{p}}=0. 
\label{Eqcondition}
\end{equation}


The conditions (\ref{Eqcondition}) determine the photon sphere radius  $ (r_{p}) $ and the critical angular momentum of the photon sphere $ (L_{p}) $. Note that the photon orbits are unstable and are determined by the condition $\partial^{2} V_{\rm eff}(r)/\partial r^{2}<0$. The second condition of Eq. (\ref{Eqcondition}) yields the following expression
\begin{eqnarray}
&& \frac{4Q^2}{3} \left(2\beta^{2}r_{p}^{4}{~} -\frac{Q^2}{5}\right)  {_{2}F_{1}}\left( \frac{1}{4},\frac{1}{2},\frac{5}{4},-\frac{Q^{2}}{\beta^{2} r_{p}^{4}}\right)
	\nonumber\\
&&+\beta^{2}r_{p}^{6}\Bigg\{ \Bigg[1+4\alpha\Bigg(\frac{2M}{r_{p}^{3}}+\frac{\Lambda}{3}-\frac{2\beta^{2}}{3}\bigg(1-\sqrt{1+\frac{Q^{2}}{\beta^{2} r_{p}^{4}}}\bigg)
	\nonumber \\
&&- \frac{4Q^{2}}{3r_{p}^{4}}{~}_{2}F_{1}\left( \frac{1}{4},\frac{1}{2},\frac{5}{4},-\frac{Q^{2}}{\beta^{2} r_{p}^{4}}\right)\Bigg) \Bigg]^{\frac{1}{2}}-\frac{3M}{r_{p}} \Bigg\} =0 \,,
\label{function-rp}
\end{eqnarray}  
which we solve numerically to obtain the radius of photon sphere. The event horizon $ (r_{e}) $ and photon sphere radius $ (r_{p}) $ are listed in Table \ref{table4}, by varying several of the black hole parameters. We note that increasing the electric charge, and the GB and BI parameters leads to a decrease in the event horizon and photon sphere radius. However, the BI parameter does not play a significant role in this case. The effect of cosmological constant is slightly  different, as increasing this parameter from $ -0.15 $ to $ -0.01 $ results in a decrease of the photon sphere radius and an increase of the event horizon. A remarkable point regarding the GB parameter is that its increasing leads to an imaginary value for the event horizon, which is physically meaningless. This shows that specific constraints should be imposed on this parameter.  
 \begin{table*}[htb!]
\centering
\caption{The event horizon and photon sphere radius for the variation of the electric charge, $Q$, the BI parameter, $\beta$, the GB parameter, $\alpha$, and the cosmological constant, $\Lambda$, for $M =1$.} \label{table4}
\begin{tabular}{c c c c c c}
 \footnotesize $Q$  \hspace{0.3cm} & \hspace{0.3cm}$0.1$ \hspace{0.3cm} &
\hspace{0.3cm} $0.3$\hspace{0.3cm} & \hspace{0.3cm} $0.6$\hspace{0.3cm} & \hspace{0.3cm}$0.9$\hspace{0.3cm} & \hspace{0.3cm} \\ \hline\hline
$ r_e (\alpha=\beta=0.2 $, $\Lambda=-0.02 $)  & $ 1.84 $ & $1.80$ & $1.63$&$1.30$ \\ 
$ r_p (\alpha=\beta=0.2 $, $\Lambda=-0.02 $) & $ 2.91 $ & $2.85$ & $2.64$&$2.18$\\\hline
\\
\footnotesize $\alpha$  \hspace{0.3cm} & \hspace{0.3cm}$0.2$ \hspace{0.3cm} &
\hspace{0.3cm} $0.5$\hspace{0.3cm} & \hspace{0.3cm} $0.8$\hspace{0.3cm} & \hspace{0.3cm}$0.9$\hspace{0.3cm} & \hspace{0.3cm} \\ \hline\hline
$ r_e (Q=\beta=0.2 $, $\Lambda=-0.02 $)  & $ 1.83 $ & $1.64$ & $1.37$&$0.98+0.2I$ \\ 
$ r_p (Q=\beta=0.2 $, $\Lambda=-0.02 $) & $ 2.89 $ & $2.73$ & $2.54$&$2.46$\\\hline
\\
\footnotesize $\beta$  \hspace{0.3cm} & \hspace{0.3cm}$0.05$ \hspace{0.3cm} &
\hspace{0.3cm} $0.1$\hspace{0.3cm} & \hspace{0.3cm} $0.15$\hspace{0.3cm} & \hspace{0.3cm}$0.25$\hspace{0.3cm} & \hspace{0.3cm} \\ \hline\hline
$ r_e (Q=\alpha=0.2 $, $\Lambda=-0.02 $) & $1.8291$  & $1.8283$ & $1.8281$ & $1.8280$\\ 
$ r_p (Q=\alpha=0.2 $, $\Lambda=-0.02 $)  & $2.8774$ & $2.8762$ & $2.8760$ & $2.8759$\\ \hline
\\
 \footnotesize $\Lambda$  \hspace{0.3cm} & \hspace{0.3cm}$-0.01$ \hspace{0.3cm} &
\hspace{0.3cm} $-0.06$\hspace{0.3cm} & \hspace{0.3cm} $-0.1$\hspace{0.3cm} & \hspace{0.3cm}$-0.15$\hspace{0.3cm} & \hspace{0.3cm}  \\ \hline\hline
$ r_e$ ($\alpha=\beta=Q=0.2$) &  $1.85$ & $1.76$ & $1.70$&$1.63$ \\
$ r_p$ ($\alpha=\beta=Q=0.2$) & $2.88$ & $2.90$ & $2.92$&$2.94$\\
 \hline
\end{tabular}
\end{table*}

\begin{figure*}[htb!]
\centering
\subfigure[~$\alpha=\beta=0.2$ and $\Lambda =-0.02$]{
   \label{Fig10a}     \includegraphics[width=0.32\textwidth]{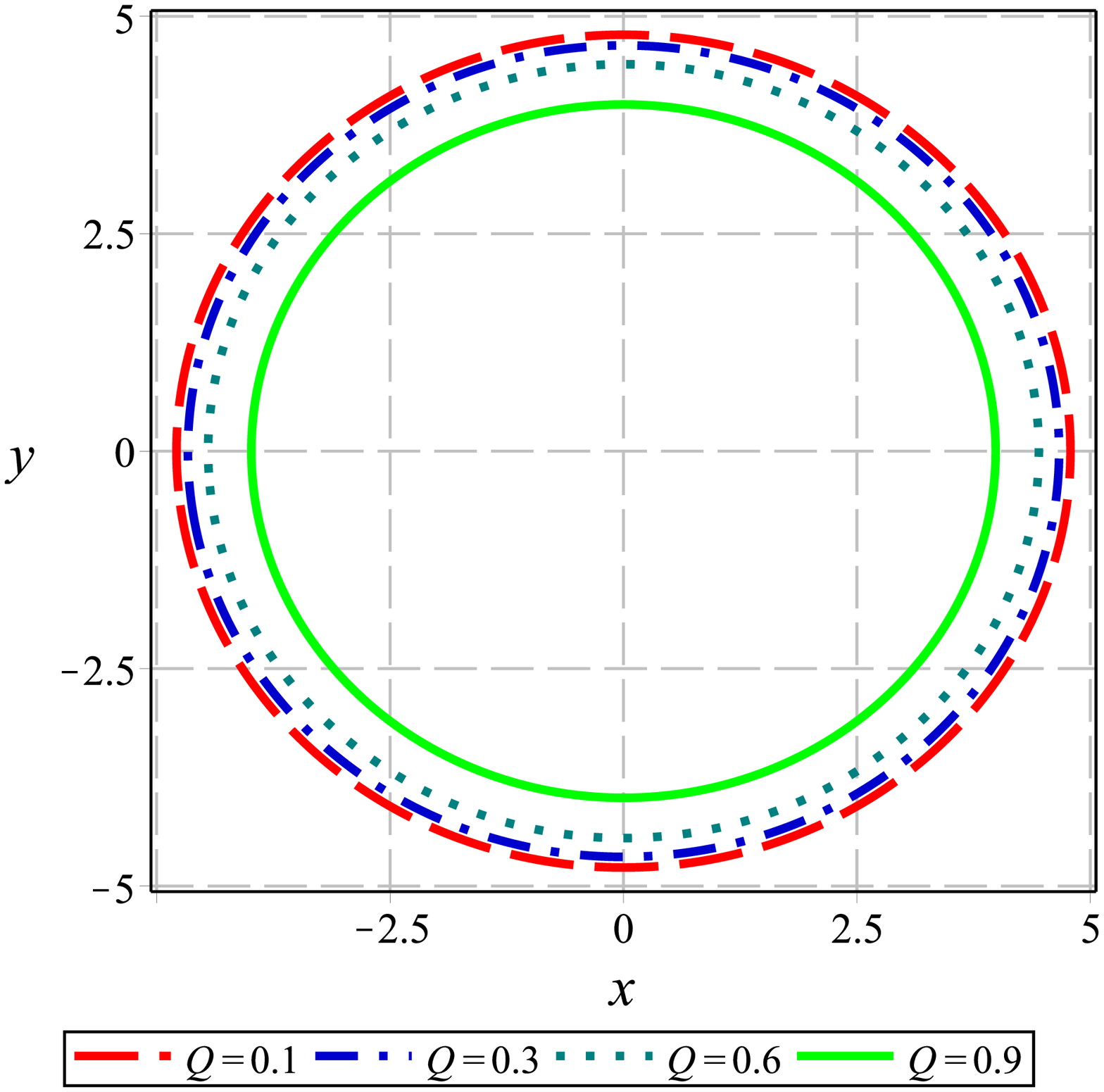}}
\subfigure[~$Q=\beta=0.2$ and $\Lambda =-0.02$]{
   \label{Fig10b}     \includegraphics[width=0.32\textwidth]{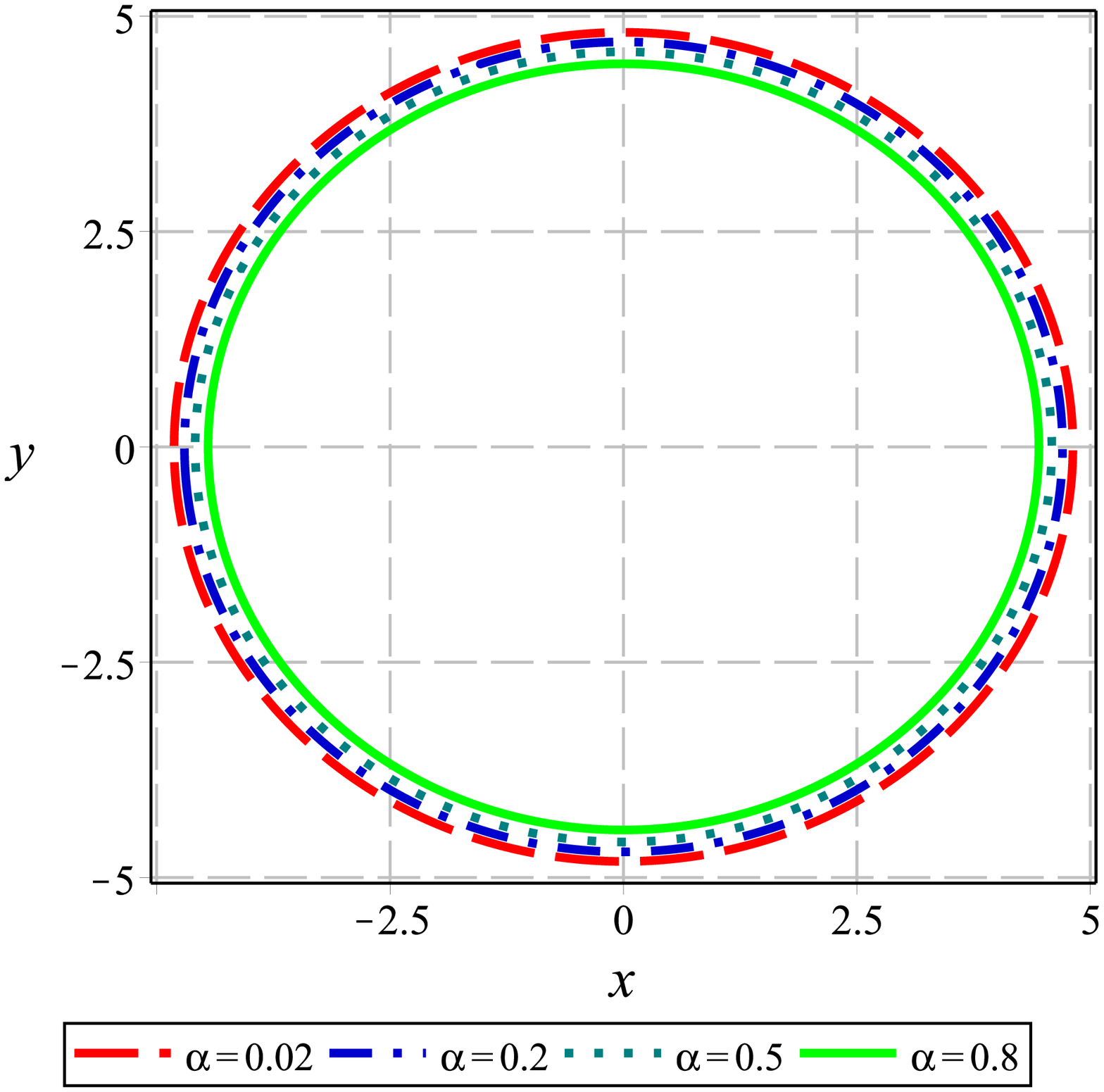}}
\subfigure[~$\alpha=\beta=Q=0.2$ ]{
     \label{Fig10c}   \includegraphics[width=0.315\textwidth]{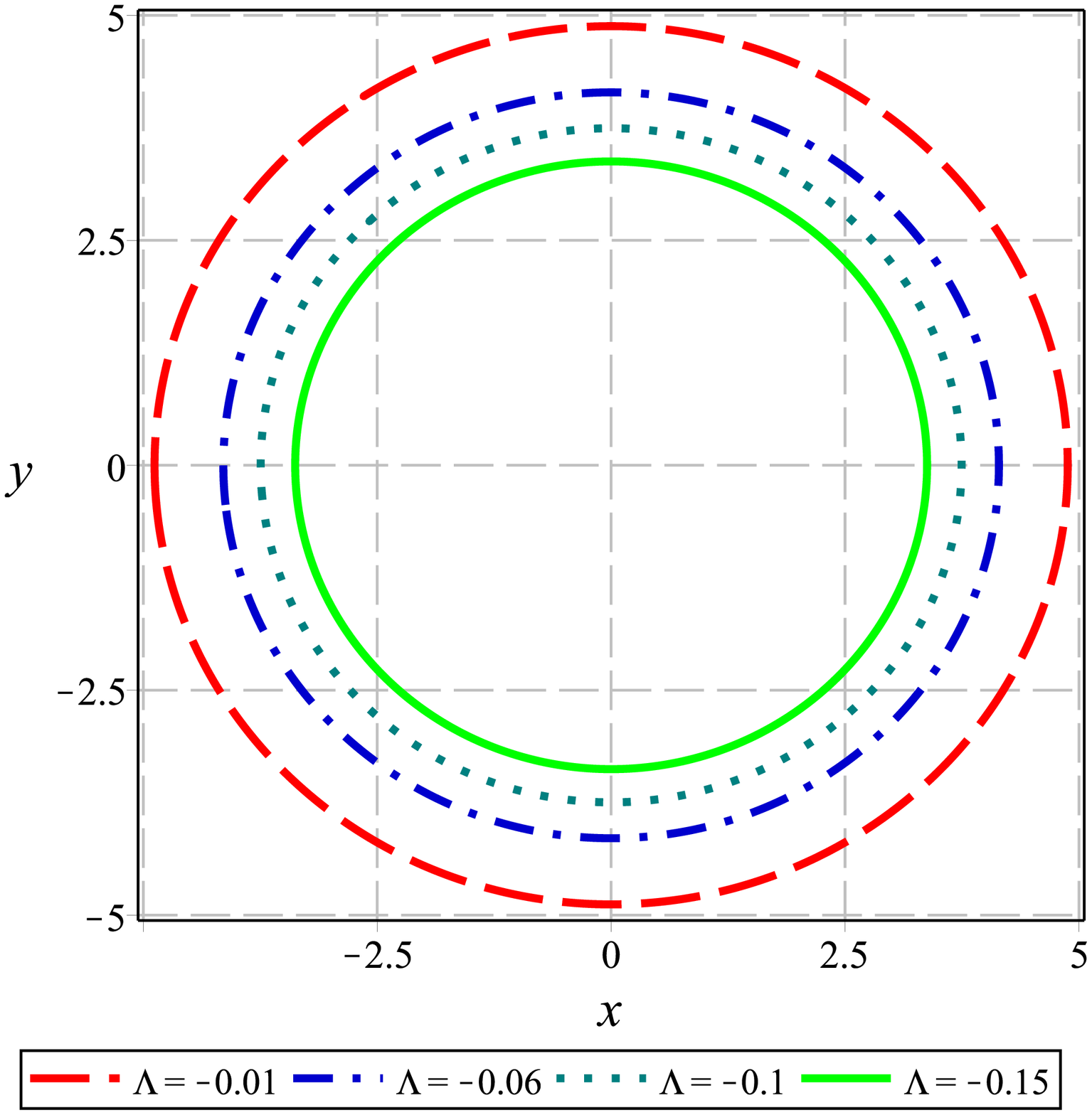}}
\subfigure[~$Q=\alpha=0.2$ and $\Lambda =-0.02$]{
      \label{Fig10d}  \includegraphics[width=0.32\textwidth]{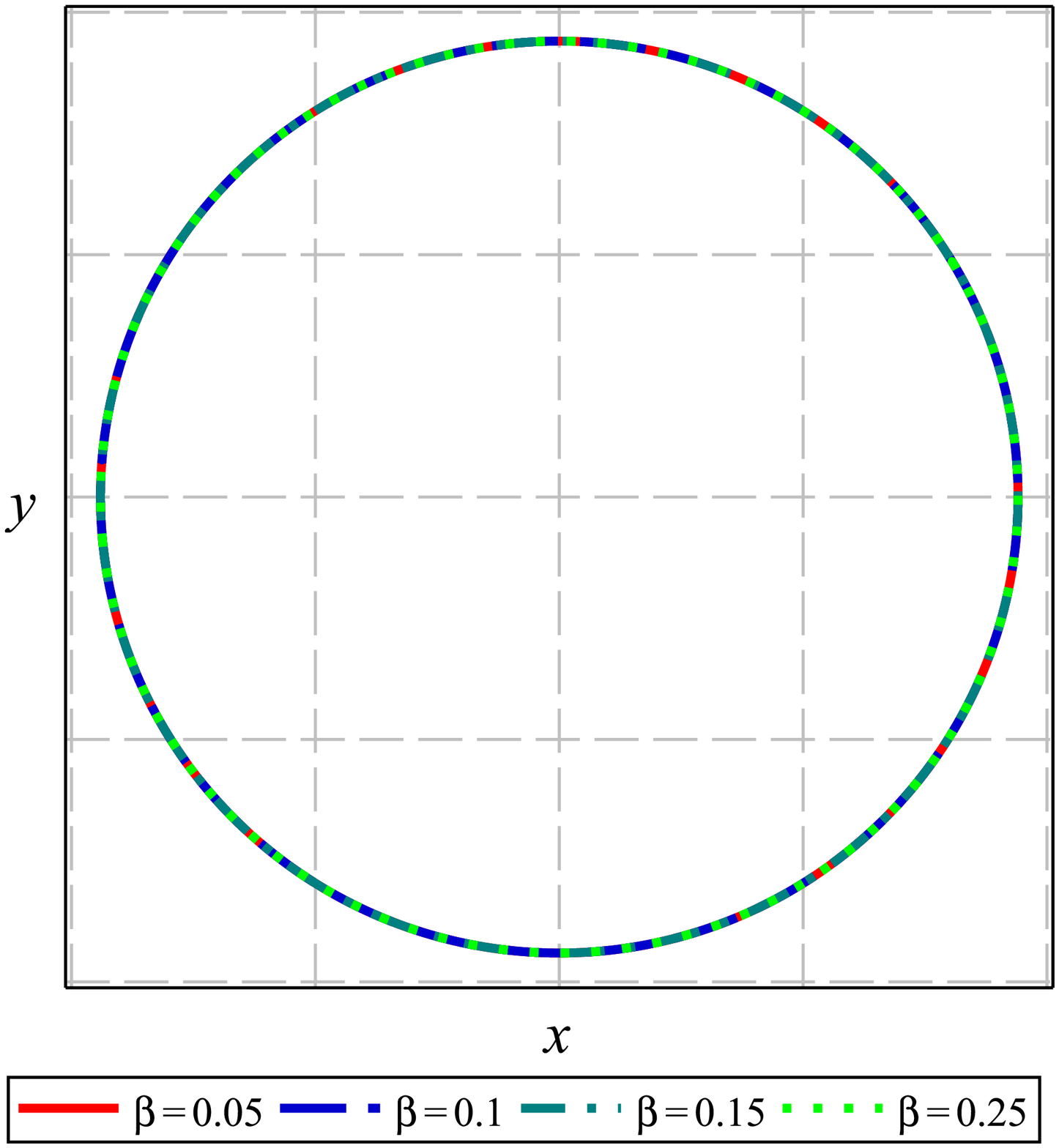}}
\subfigure[~$Q=\alpha=0.2$ and $\Lambda =-0.02$]{
    \label{Fig10e}    \includegraphics[width=0.33\textwidth]{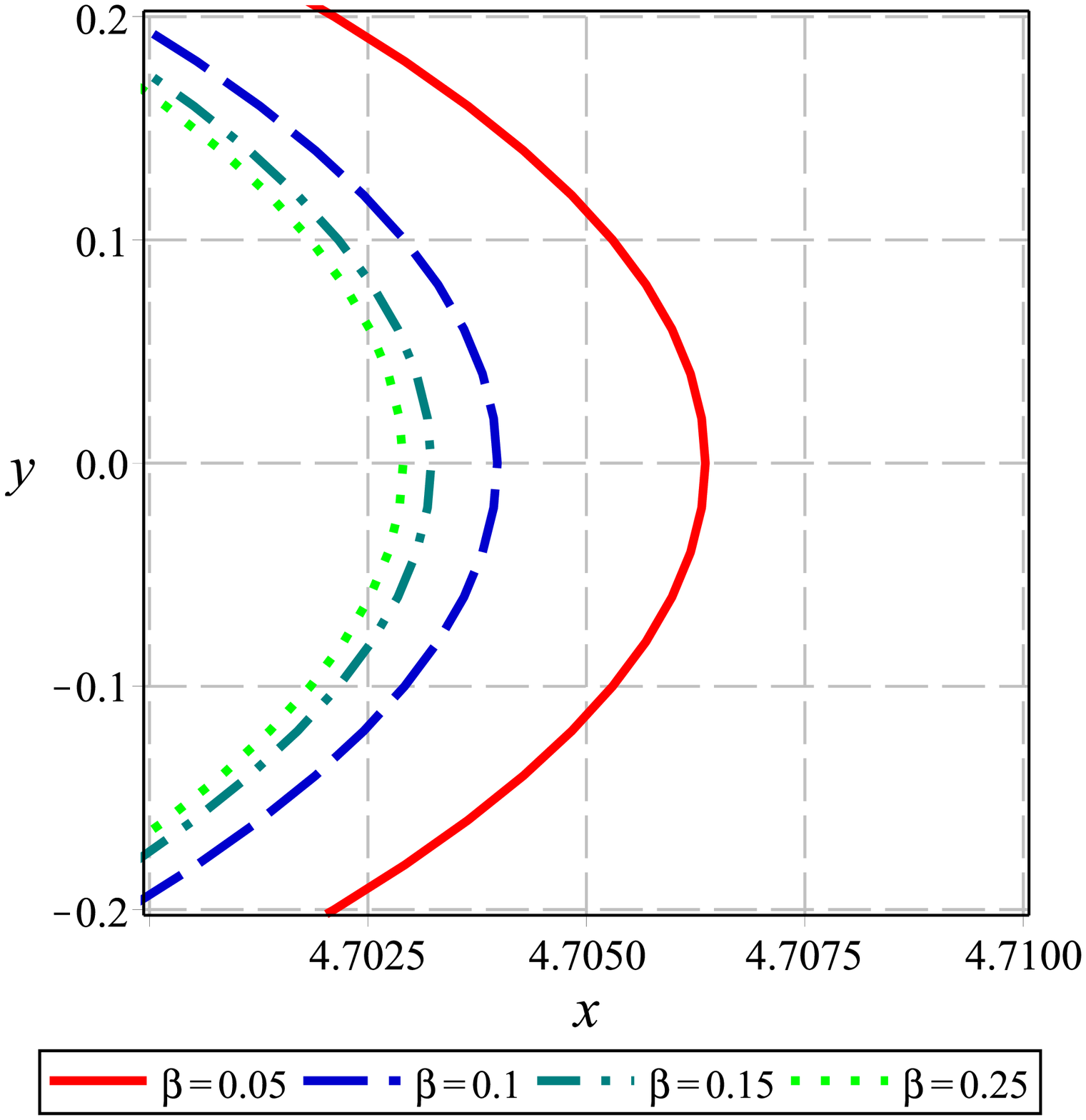}}
\caption{The plots depict the black hole shadow in the celestial plane $(x -y )$ for $M=1$. See the text for more details.}
\label{Fig10}
\end{figure*}

As a next step in the analysis, we examine the behavior of the shadow radius in the corresponding NLED theory  coupled to EGB gravity. From the definition of the shadow radius \cite{Perlick1}, the size of black hole shadow can be expressed in celestial coordinates $(x,y)$ as
\begin{equation}
 r_{s}=\sqrt{x^{2}+y^{2}}=\frac{L_{p}}{E}=\frac{r_{p}}{\sqrt{f(r_{p})}}.
\label{EqRs}
\end{equation} 
Thus, using Eq. (\ref{EqRs}) and the listed values of Table \ref{table4}, we explore the shadow size of this solution. To inspect the influence of the black hole parameters on the shadow geometric circular shape, we plot Fig. \ref{Fig10}. Evidently,  the electric charge decreases the shadow size for fixed values of $  \alpha$, $ \beta $ and $ \Lambda $ (see Fig. \ref{Fig10a}). Taking into account Figs. \ref{Fig10b} and \ref{Fig10e}, we verify that the GB and BI parameters have a decreasing effect on the shadow radius similar to the electric charge. However, the variation of the BI parameter $ \beta $ does not affect the shadow size significantly. Figure \ref{Fig10c} depicts the impact of the cosmological constant on the size of the shadow. As we see, unlike the previous parameters, the cosmological constant increases the shadow radius and has a notable effect on it. 


It has been known that the black hole shadow corresponds to its high energy absorption cross section for a distant observer. Indeed, the absorption cross section oscillates near to a limiting constant value $ \sigma_{\rm lim} $ for a spherically symmetric black hole. According to \cite{WWei,Belhaj1,Belhaj2}, $ \sigma_{\rm lim} $ was found to be equal to the area of the photon sphere ($ \sigma_{\rm lim}\approx \pi r_{s}^{2} $). Since the shadow measures the optical appearance of a black hole, it can be equal to the limiting constant value of the high-energy absorption cross section.  The energy emission rate is given by
\begin{equation}
\frac{d^{2}E(\omega)}{dtd\omega}=\frac{2\pi^{3}\omega^{3}r_{s}^{2}}{e^{\frac{\omega}{T}} -1},
\label{Eqemission}
\end{equation}
where $ \omega $ and $ T $ are, respectively, the emission frequency and Hawking temperature. For the present solution, the Hawking temperature is obtained as
\begin{equation}
T=\frac{1}{4\pi r_{e}}\left[\frac{-\Lambda r_{e}^{4}-3\alpha +2\beta^{2}r_{e}^{4}\left( 1-\sqrt{1+\frac{Q^{2}}{\beta^{2}r_{e}^{4}}}\right) }{r_{e}^{2}+2\alpha} +1 \right] .
\label{EqTH1}
\end{equation}

\begin{figure*}[htb!]
\centering
\subfigure[~$\alpha=\beta=0.2$ and $\Lambda =-0.02$]{
   \label{Fig11a}     \includegraphics[width=0.32\textwidth]{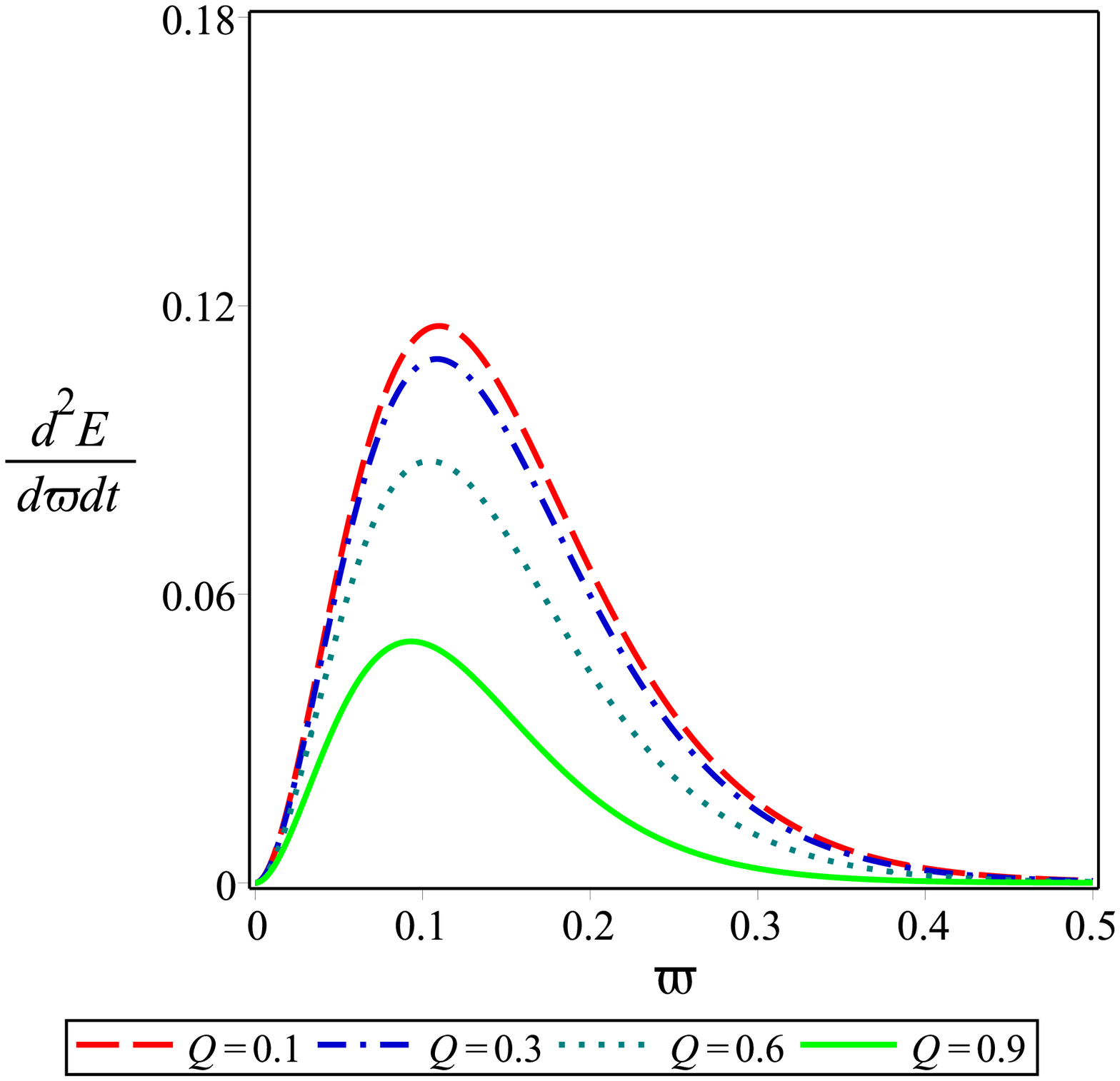}}
    \hspace{1cm}
\subfigure[~$Q=\beta=0.2$ and $\Lambda =-0.02$]{
   \label{Fig11b}     \includegraphics[width=0.32\textwidth]{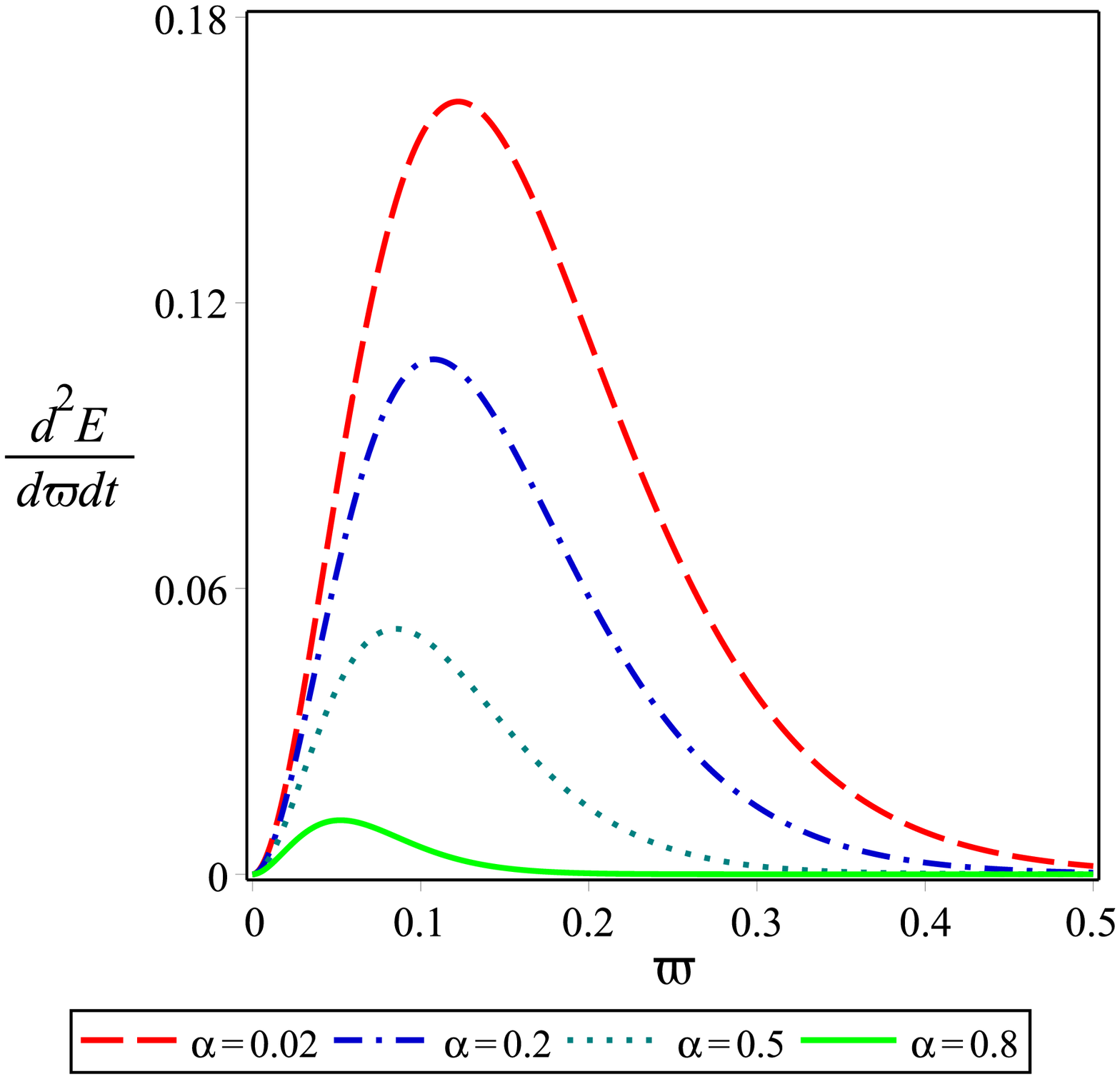}}
\\    
\subfigure[~$Q=\alpha=0.2$ and $\Lambda =-0.02$ ]{
      \label{Fig11c}  \includegraphics[width=0.32\textwidth]{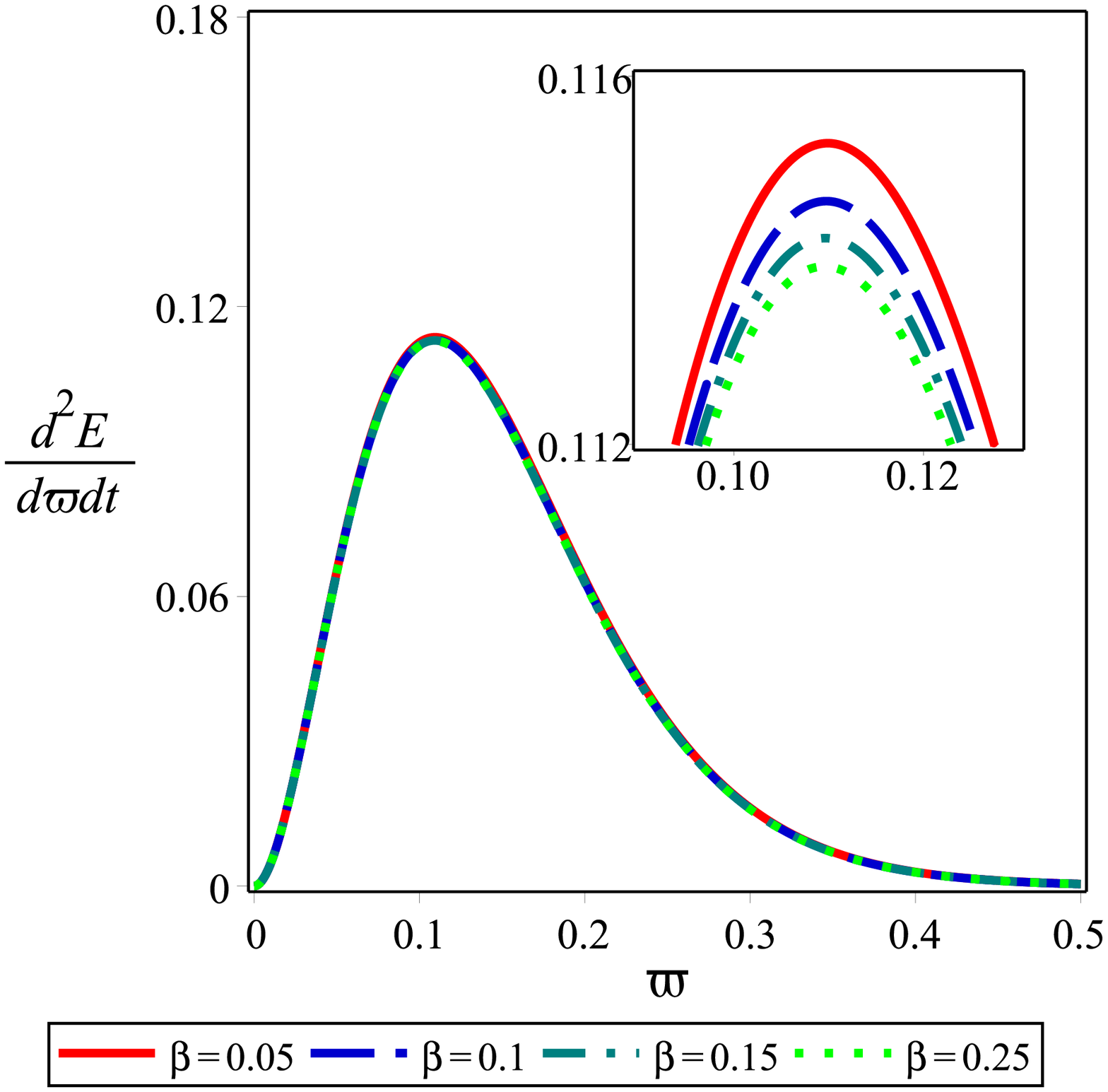}}
       \hspace{1cm}
\subfigure[~$\alpha=\beta=Q=0.2$]{
      \label{Fig11d}  \includegraphics[width=0.32\textwidth]{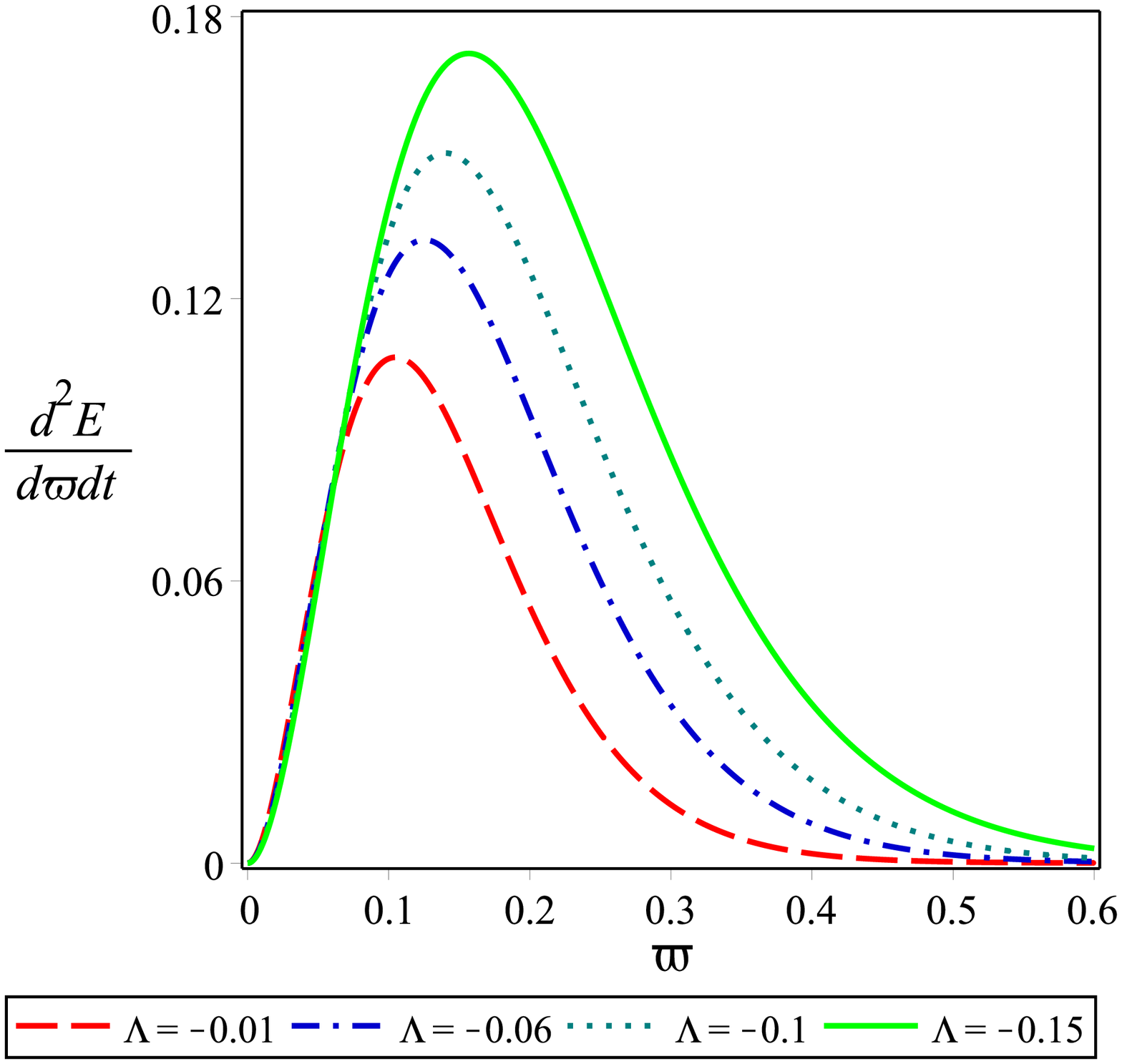}}
\caption{The energy emission rates for the corresponding black hole, with $ M=1 $, and different values of $ Q $, $ \beta $, $ \alpha $ and $  \Lambda$. See the text for for details.}
\label{Fig11}
\end{figure*}

Figure \ref{Fig11} depicts the behavior of the BH parameters on the energy emission. As it is transparent, there exists a peak of the energy emission rate for the black hole. As the BH parameters increase, the peak decreases and shifts to the low frequency. From Fig. \ref{Fig11a}, we verify that the electric charge decreases the energy emission, so that when the black hole is located in a more powerful electric field, the evaporation process would be slower.
Regarding the effects of the GB and BI parameters, we see from  Figs. \ref{Fig11b} and \ref{Fig11c}, that increasing these two parameters results in the decrease of the energy emission. In fact,  decreasing the coupling constants implies a fast emission of particles. But as in the shadow case, the effect of $ \beta $ is negligible. Figure \ref{Fig11d} displays the decreasing contribution of the cosmological constant on the emission rate which shows that the black hole should be located in a lower curvature background in order to have a faster evaporation. The figure also shows that  the variation of  cosmological constant has a stronger effect on the radiation rate than other parameters.

\section{Connection between shadow radius and Quasinormal modes}\label{sec:IV}

Quasinormal modes (QNMs) are characteristic frequencies with a nonvanishing imaginary part, which encode important information related to the stability of the black hole under small perturbations. These frequencies depend on the details of the geometry and the type of the perturbation, but not on the initial conditions.
To study these modes, an outgoing boundary condition should be imposed  at infinity and an ingoing boundary condition at the horizon. In general, QNMs are characterized by complex numbers, $ \omega = \omega_{R} +i \omega_{I}$. The sign of the imaginary part determines if the mode is stable or unstable. For positive  $ \omega_{I} $ (exponential growth), the mode is unstable, whereas negativity represents stable modes. In fact, it was shown that the real and imaginary parts of QNMs in the eikonal limit are, respectively, related to the angular velocity and Lyapunov exponent of unstable circular null geodesics \cite{QNM1}. It is worthwhile to mention that such a correspondence is only guaranteed for test fields, and not for gravitational ones \cite{QNM22}. Such a connection was investigated between QNMs and gravitational lensing \cite{QNM2}.

Recently, it has been suggested that the real part of QNMs in the eikonal limit can be connected to the radius of the black hole shadow \cite{QNM3,QNM4}. Such a correspondence has been applied to different black holes \cite{QNM5,QNM6,QNM7}. Here, we employ this idea and study the QNMs for the black hole solution presented by the metric function given in Eq. (\ref{function-a}). According to this correspondence, the quasinormal frequency $ \omega $ can be calculated with the property of the photon sphere as 
\begin{equation}
\omega = l \Omega -i \left(n+\frac{1}{2} \right) | \lambda |.
\label{EqQNM1}
\end{equation}
where $ n $ and $ l $ are  the overtone number {and the multiple number (also called the angular quantum number)}, respectively.  The quantities $ \Omega $ and $ \lambda $ are the angular velocity and Lyapunov exponent of the photon sphere, which can be obtained as
\begin{equation}
\Omega = \frac{\sqrt{f(r_{p})}}{r_{p}}=\frac{1}{L_{p}},\label{EqQNM2}
\end{equation}
\begin{equation}
\lambda = \frac{\sqrt{2f(r_{p})-r_{p}^{2}f^{\prime \prime}}}{\sqrt{2}L_{p}}.
\label{EqQNM3}
\end{equation}
\begin{figure*}[htb!]
\centering
\subfigure[~$\beta=0.2$ and $\Lambda =-0.02$]{
  \label{FigBQa}      \includegraphics[width=0.30\textwidth]{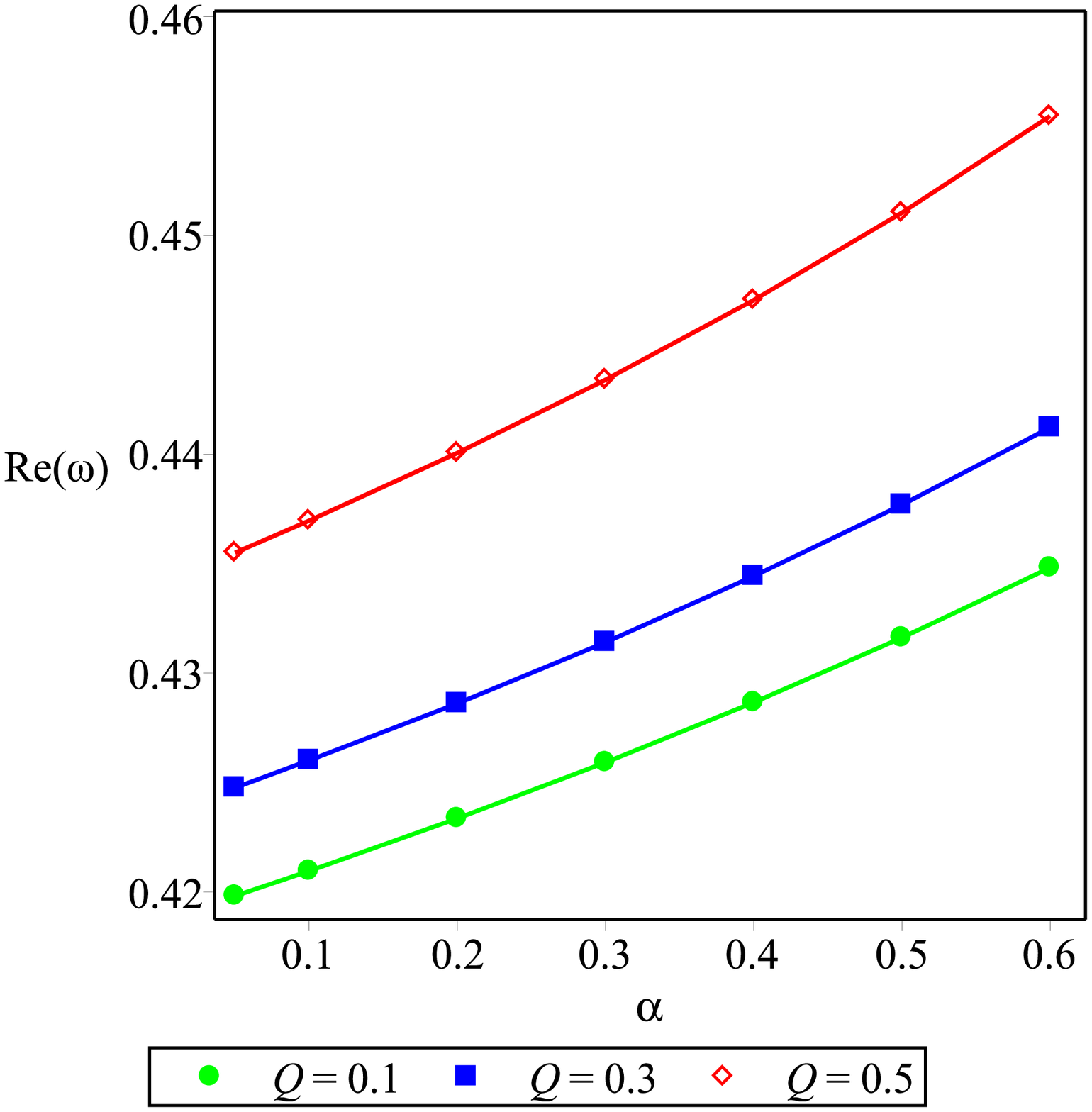}}
  \hspace{1cm}
\subfigure[~$\beta=0.2$ and $\Lambda =-0.02$]{
  \label{FigBQb}      \includegraphics[width=0.31\textwidth]{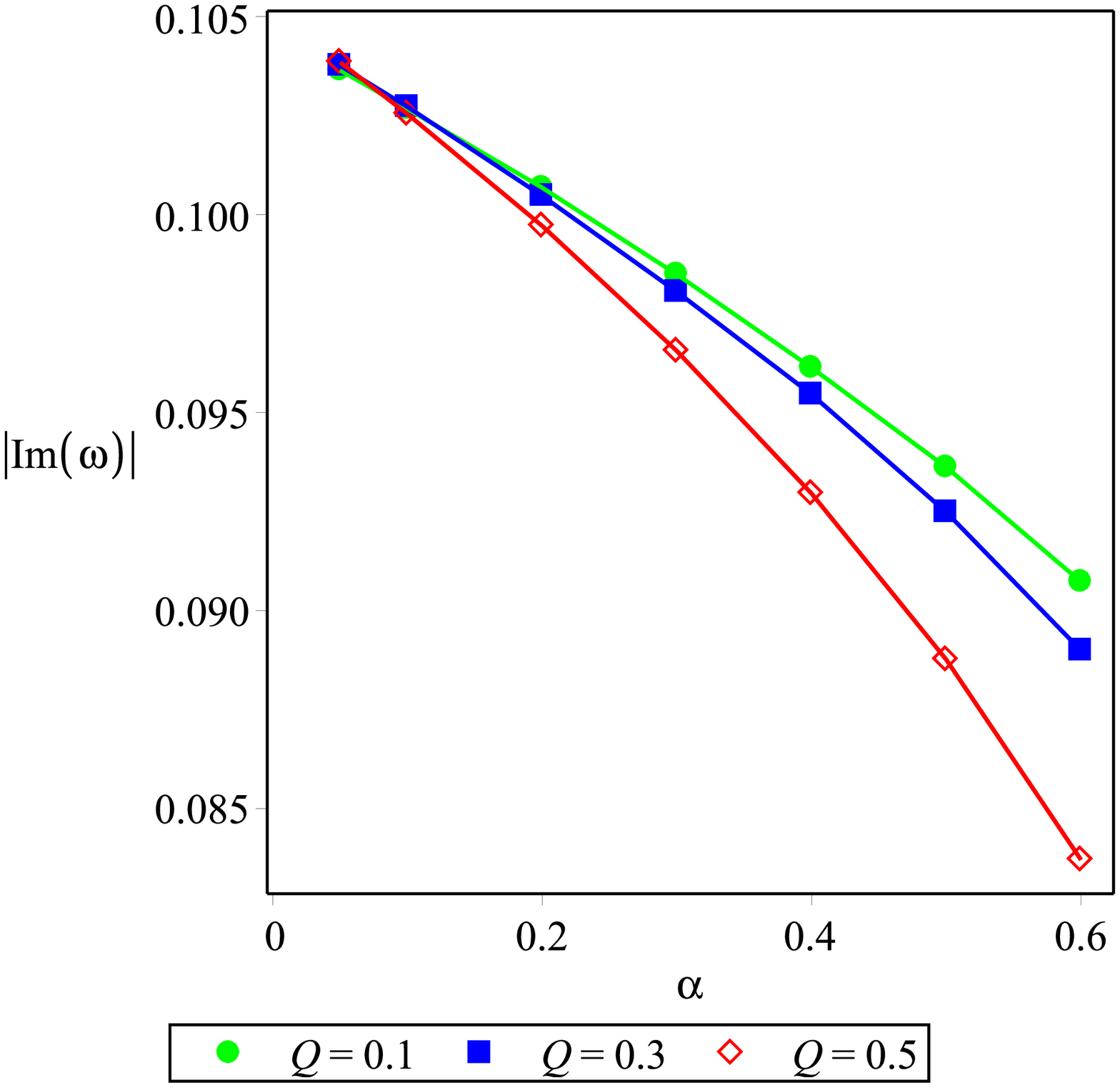}}
\\
\subfigure[~$Q=0.2$ and $\Lambda =-0.02$]{
     \label{FigBQc}   \includegraphics[width=0.30\textwidth]{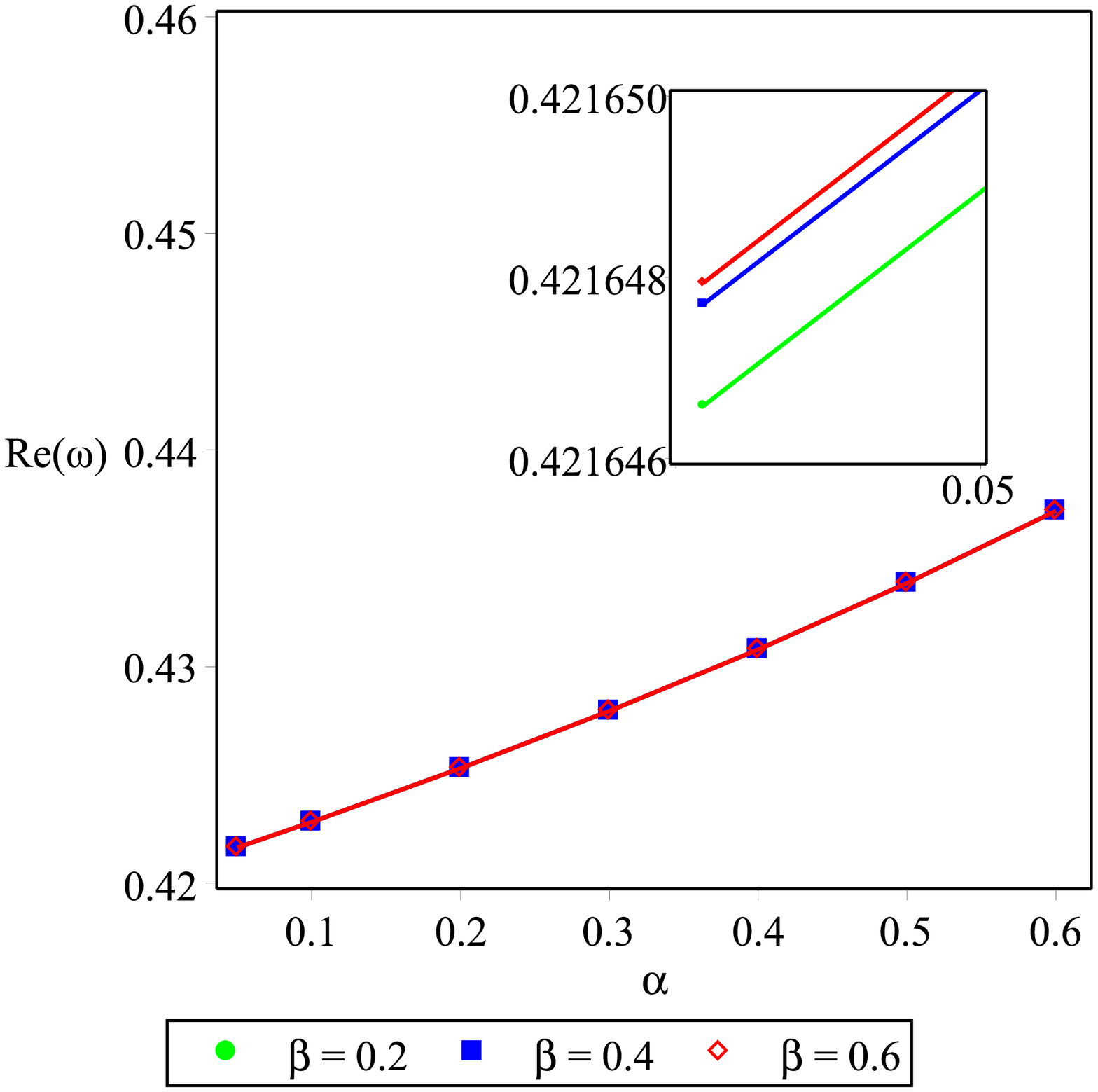}}   
     \hspace{1cm} 
\subfigure[~$Q=0.2$ and $\Lambda =-0.02$]{
    \label{FigBQd}    \includegraphics[width=0.30\textwidth]{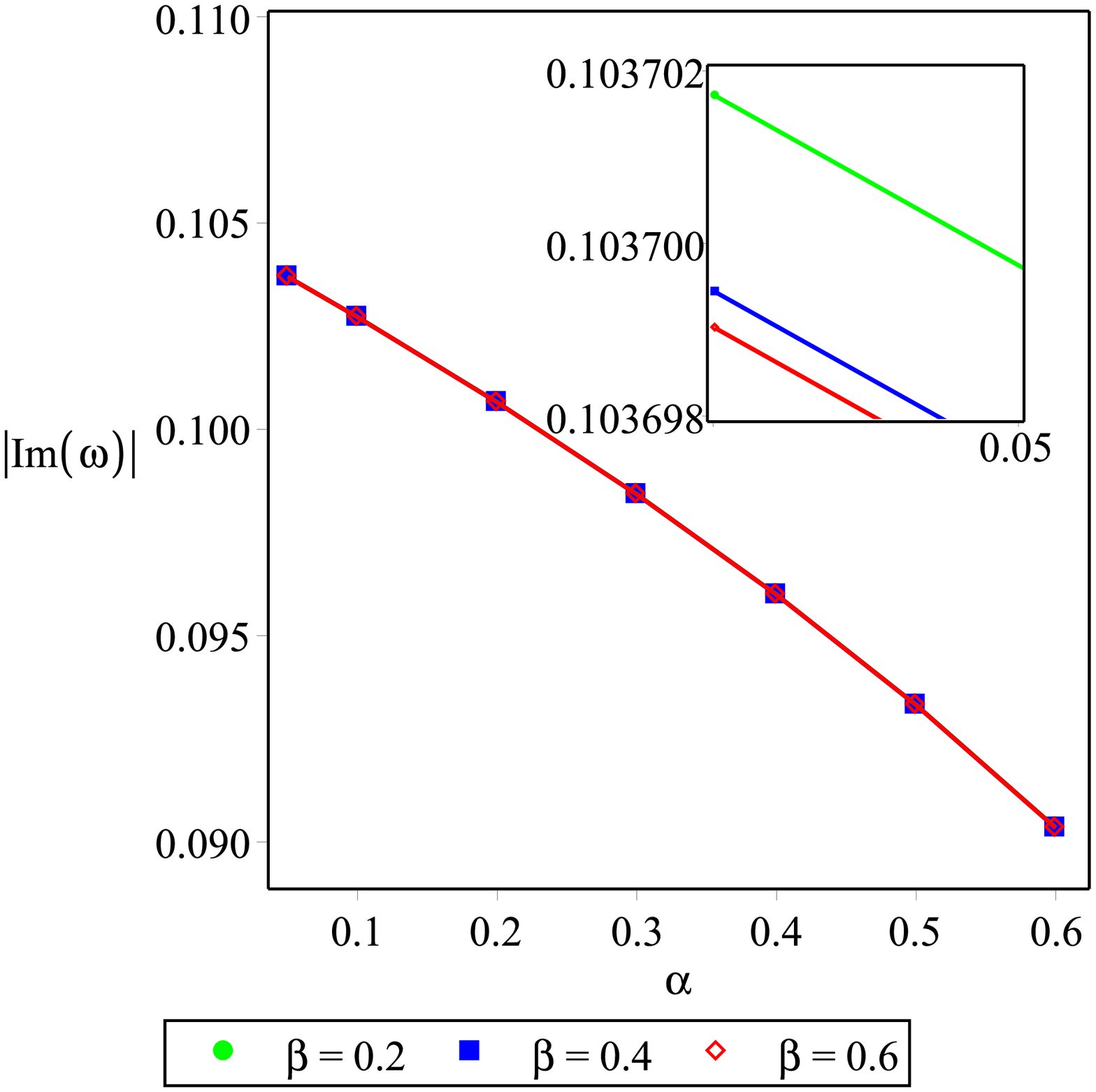}}
\\ 
\subfigure[~$Q=\beta=0.2$]{
    \label{FigBQe}    \includegraphics[width=0.30\textwidth]{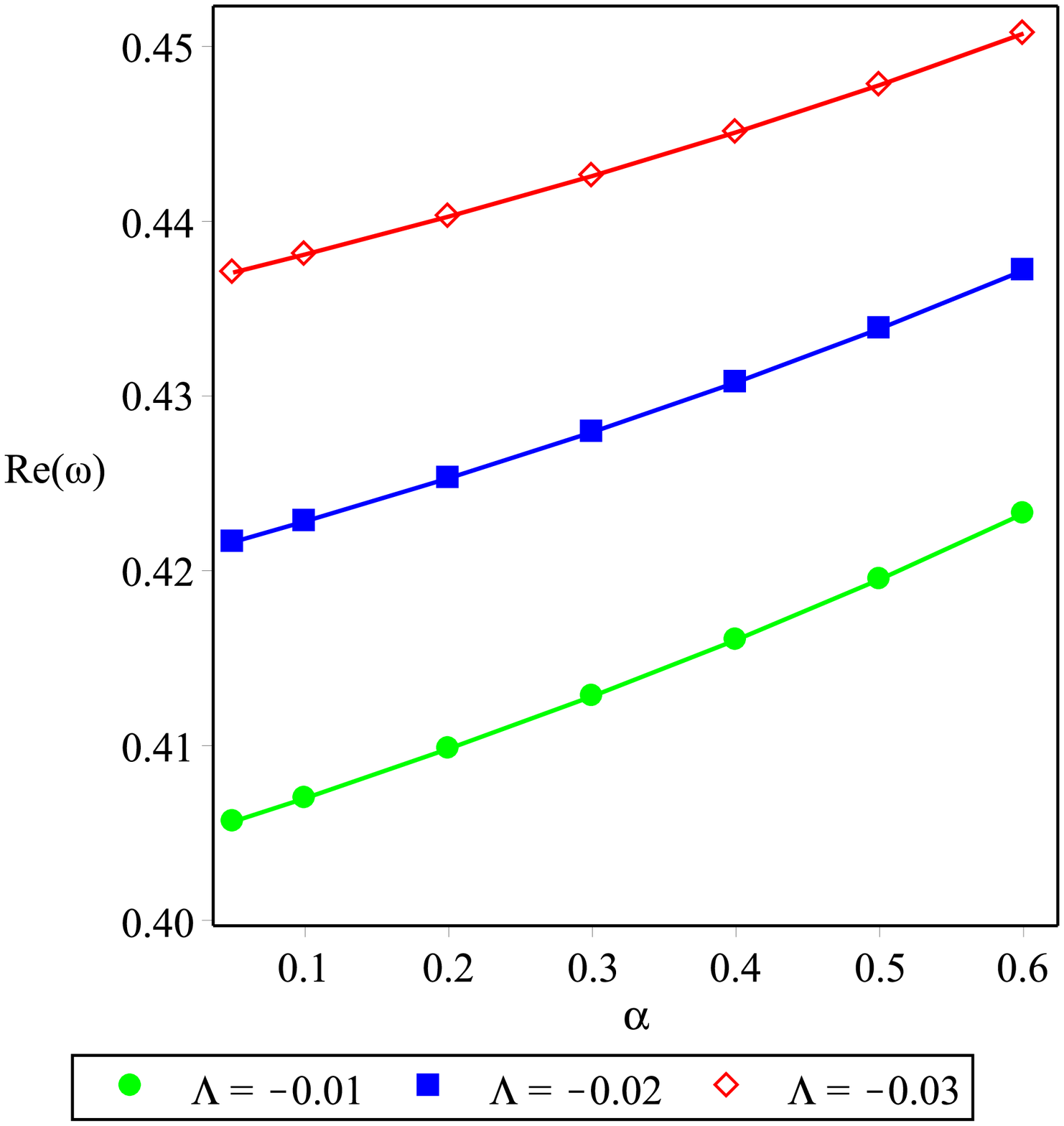}}
    \hspace{1cm}
\subfigure[~$Q=\beta=0.2$ ]{
 \label{FigBQf}       \includegraphics[width=0.32\textwidth]{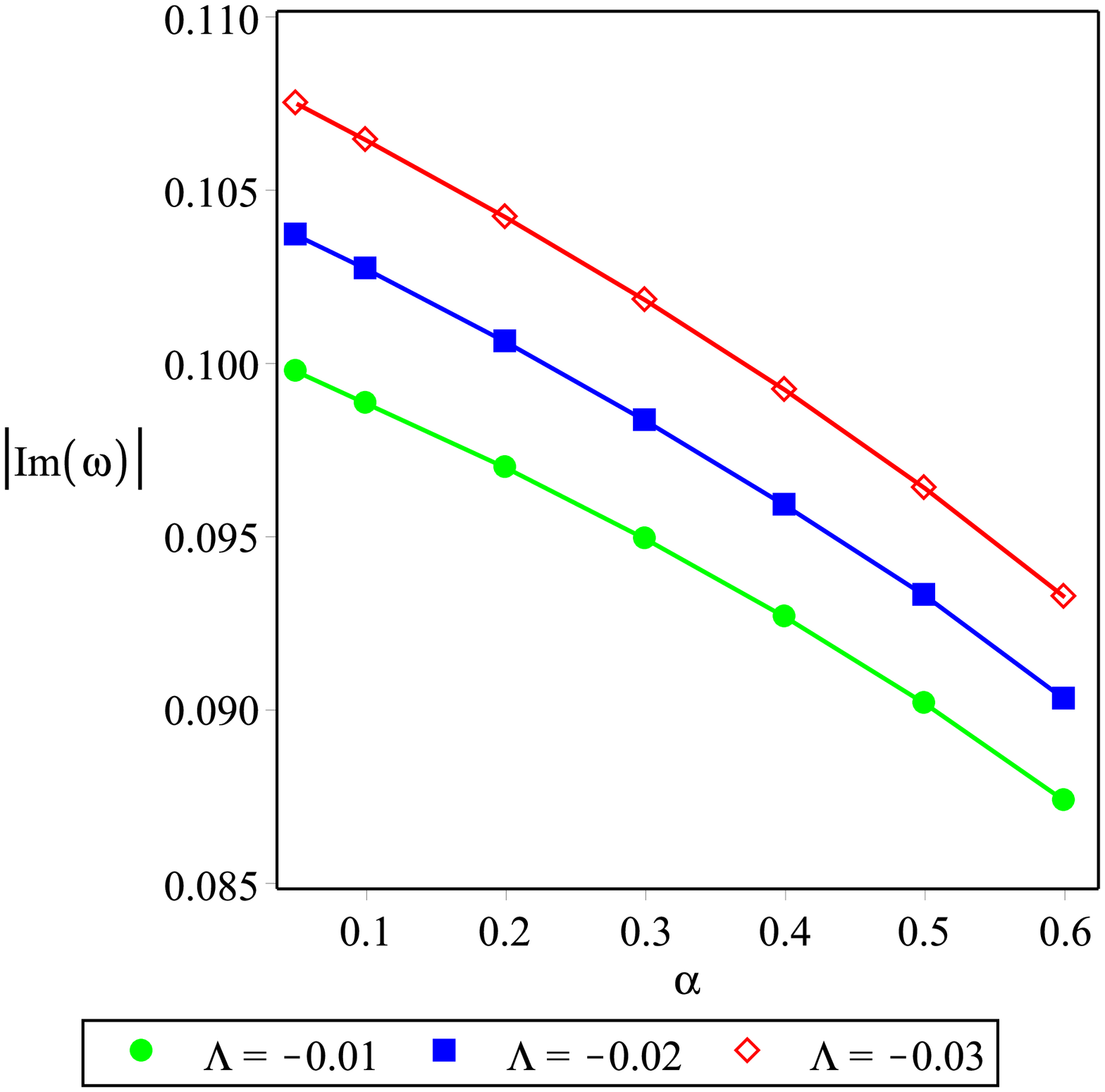}}
\caption{The behavior of $Re (\omega)$ and $Im (\omega)$ with respect to the GB parameter for $ M=1 $, $ n=0 $, $l=2$ and different values of $ Q $ (Figs. \ref{FigBQa} and \ref{FigBQb}), $ \beta $ (Figs. \ref{FigBQc} and \ref{FigBQd}) and  $ \Lambda$ (Figs. \ref{FigBQe} and \ref{FigBQf}). See the text for more details.}
\label{FigBQ}
\end{figure*}

From Eq. (\ref{EqQNM1}), evidently the real part of the modes is proportional to the angular degree, whereas the imaginary part depends on  the overtone number only. Using  Eqs. (\ref{EqQNM1}),  (\ref{EqQNM2}) and (\ref{EqQNM3}), we can investigate how the black hole parameters affect quasinormal frequencies. These effects are depicted in Fig. \ref{FigBQ}, where the spectrum obtained exhibits the following features:

\begin{itemize}
    
	\item The real (imaginary) part of the quasinormal frequencies exhibits an increasing (decreasing) behavior for increasing $\alpha$. This shows that the scalar field perturbations around a black hole with a stronger GB coupling oscillates with more energy and decays slower. 
        
    \item According to Figs. \ref{FigBQa} and \ref{FigBQb}, the electric charge increases (decreases) the real (imaginary) value of the QNM frequency. This indicates that the scalar field perturbations in the presence of electric charge oscillate faster and decay slower as compared to neutral black holes.
    
    \item From Figs. \ref{FigBQc} and \ref{FigBQd}, we verify that the effect of the BI parameter is similar to the GB parameter, but unlike the GB coupling constant, its effect is negligible.
    
    \item Both $ Re(\omega) $ and $Im(\omega)$ decrease by increasing $ \Lambda $, which is transparent from Figs. \ref{FigBQe} and \ref{FigBQf}. This implies that the existence of a higher curvature background restrains the oscillation amplitude of the scalar field, even though it could decay slower.

\end{itemize}

\begin{figure*}[htb!]
\centering
\subfigure[~$\alpha=\beta=0.2$ and $\Lambda =-0.02$]{
 \label{Fig12a}       \includegraphics[width=0.22\textwidth]{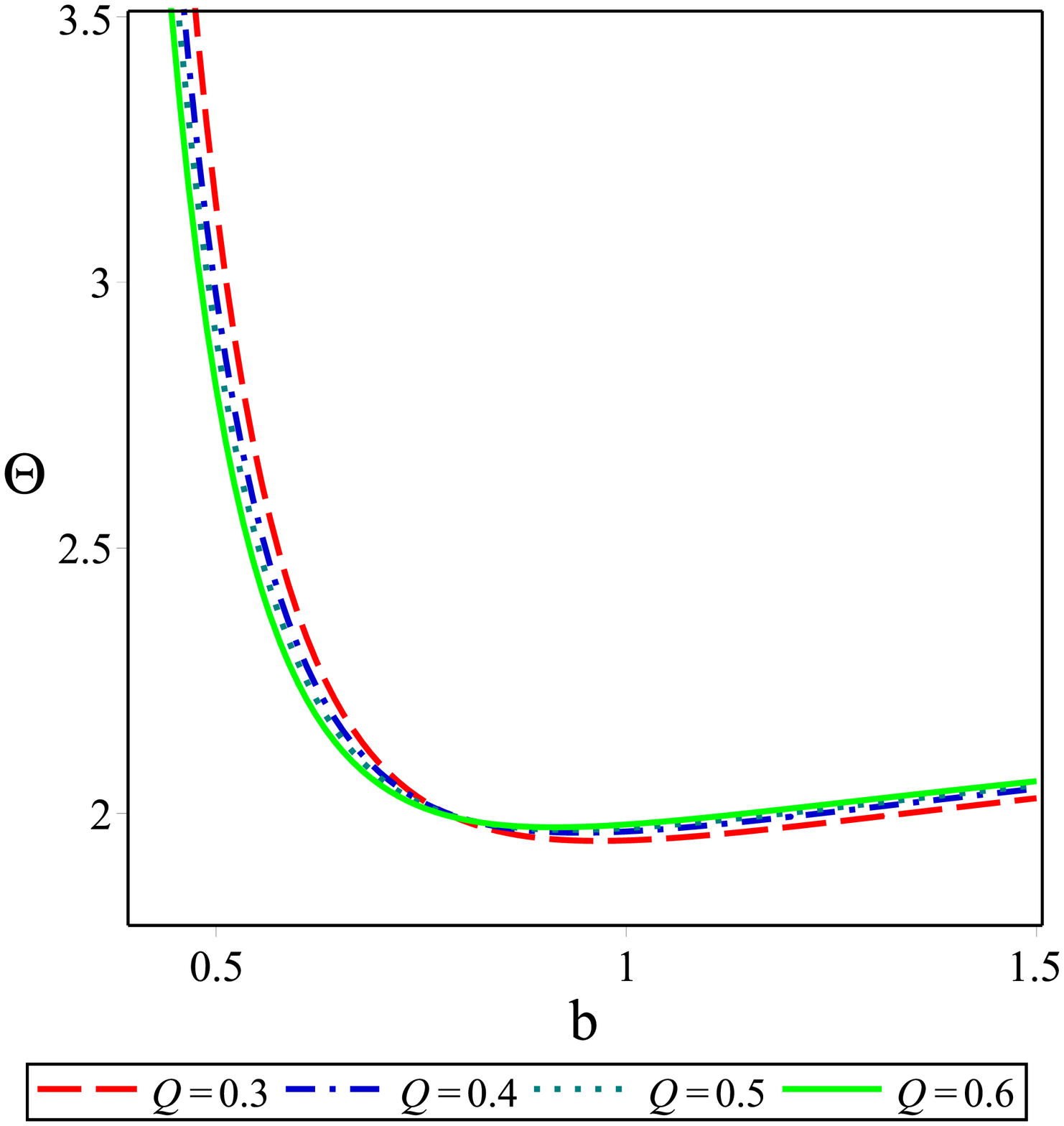}}
\subfigure[~$\alpha=\beta=0.2$ and $\Lambda =-0.02$]{
 \label{Fig12b}       \includegraphics[width=0.22\textwidth]{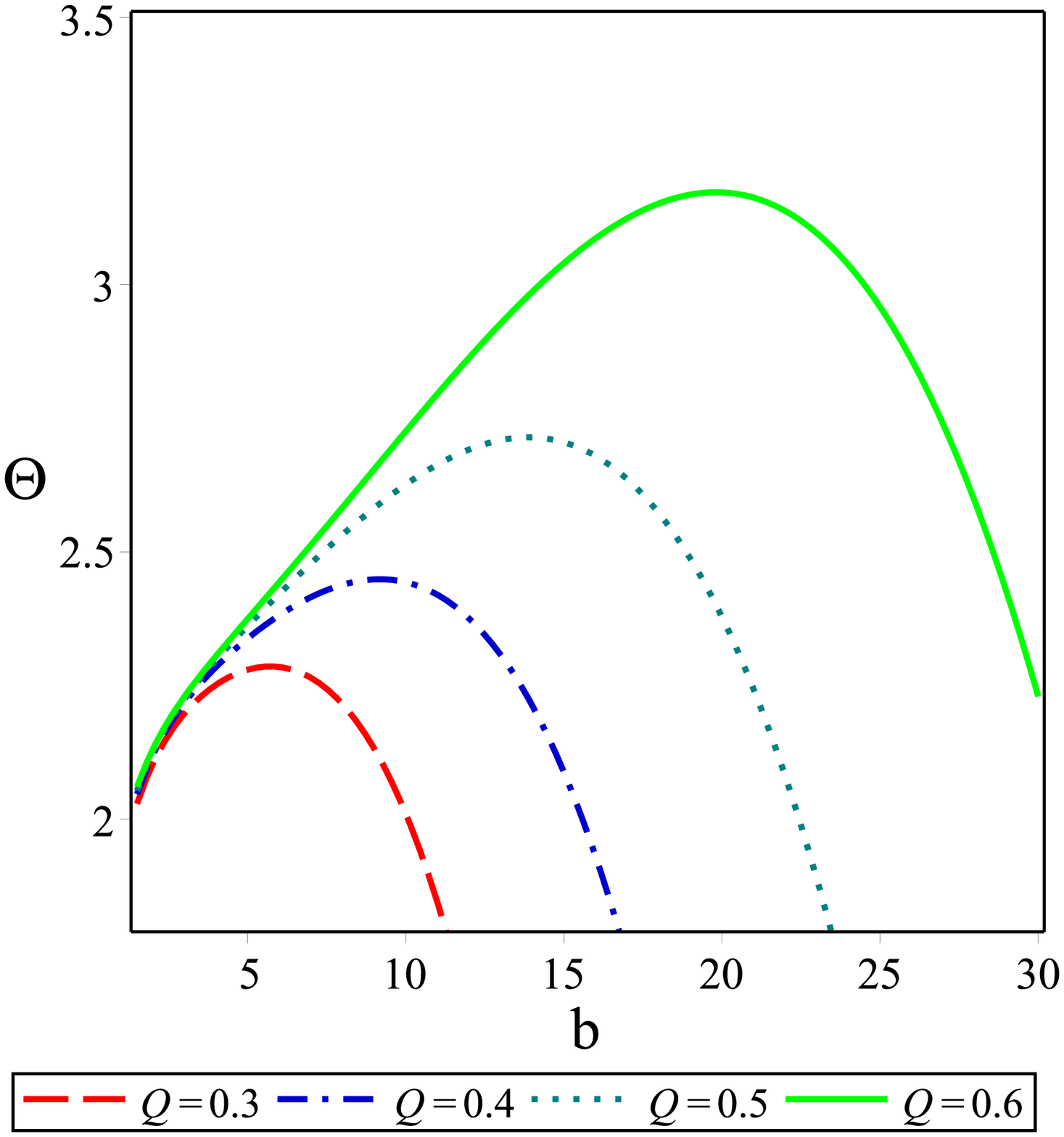}}
\subfigure[~$Q=\beta=0.2$ and $\Lambda =-0.02$]{
 \label{Fig12c}       \includegraphics[width=0.22\textwidth]{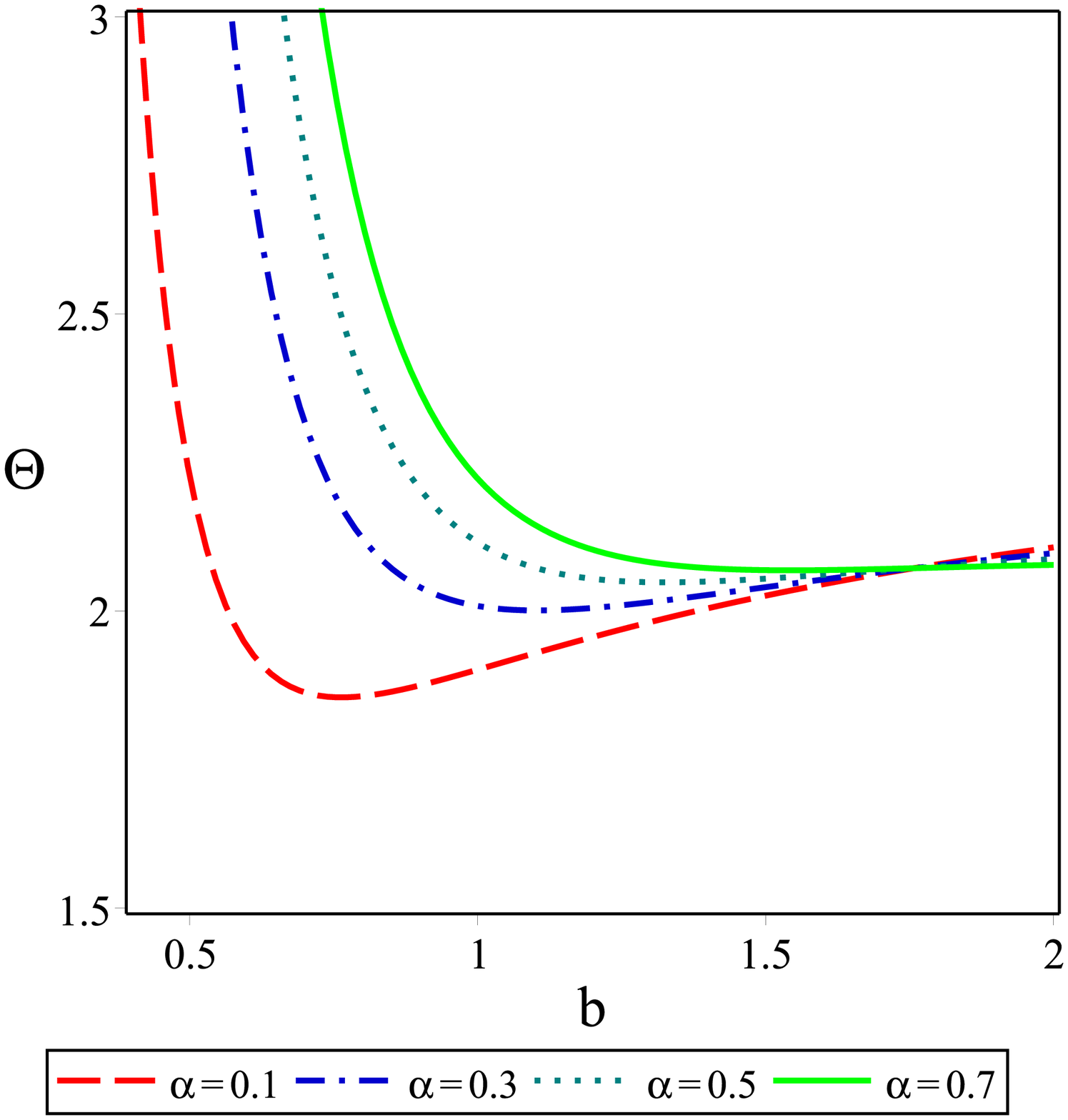}}
\subfigure[~$Q=\beta=0.2$ and $\Lambda =-0.02$]{
    \label{Fig12d}    \includegraphics[width=0.22\textwidth]{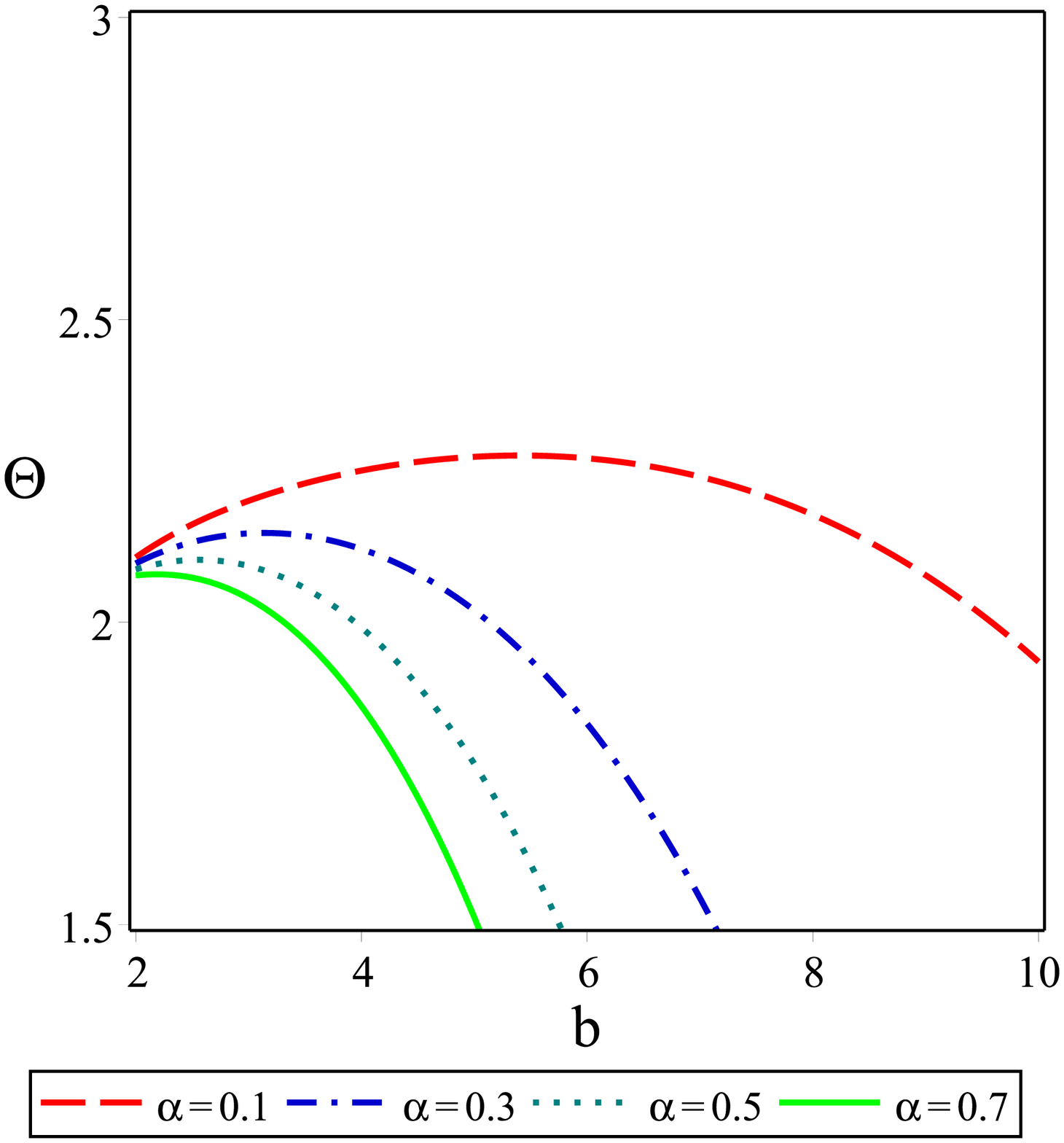}}
\\    
\subfigure[~$Q=\alpha=0.2$ and $\Lambda =-0.02$ ]{
 \label{Fig12e}       \includegraphics[width=0.22\textwidth]{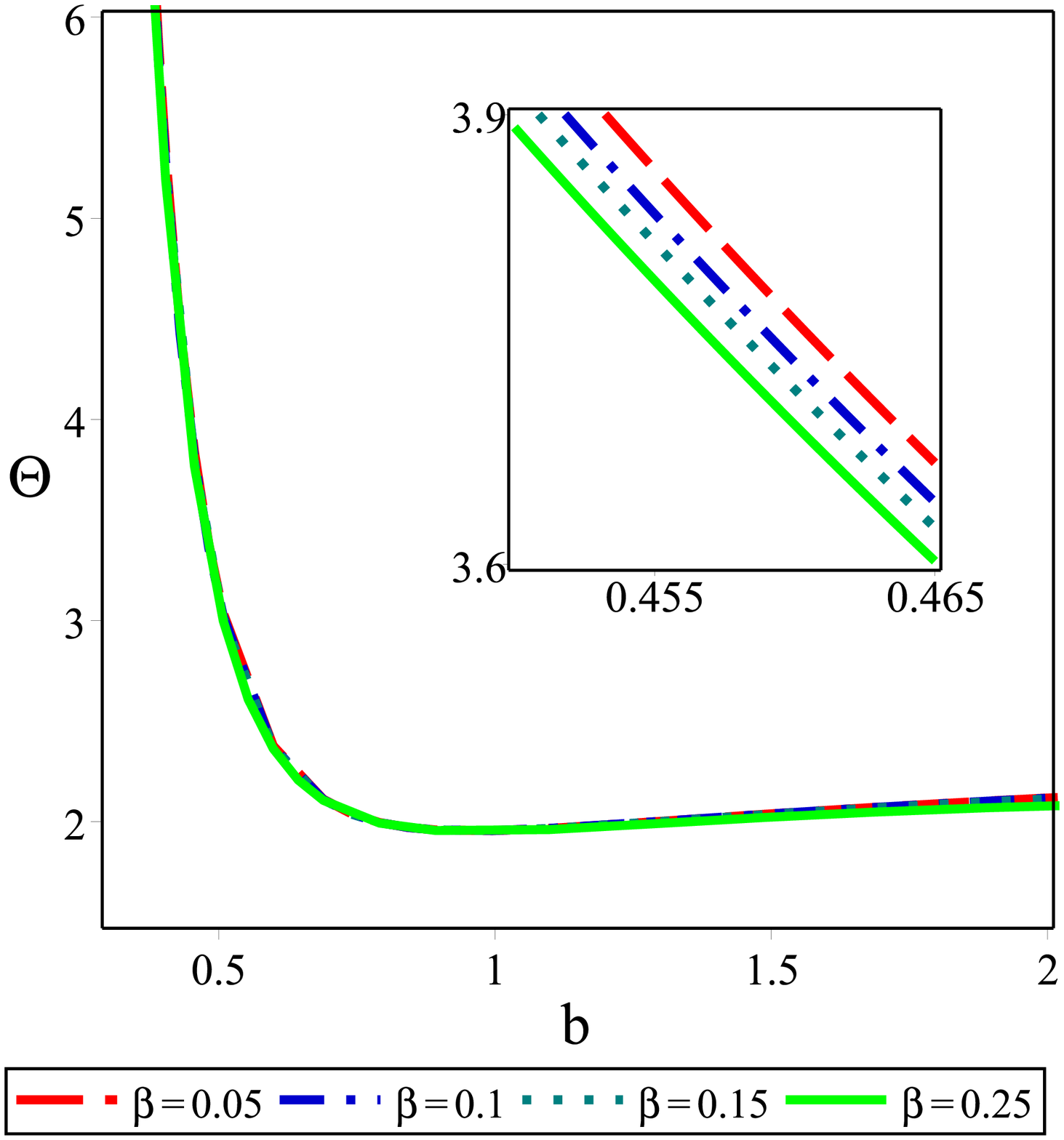}}
\subfigure[~$Q=\alpha=0.2$ and $\Lambda =-0.02$ ]{
 \label{Fig12f}       \includegraphics[width=0.22\textwidth]{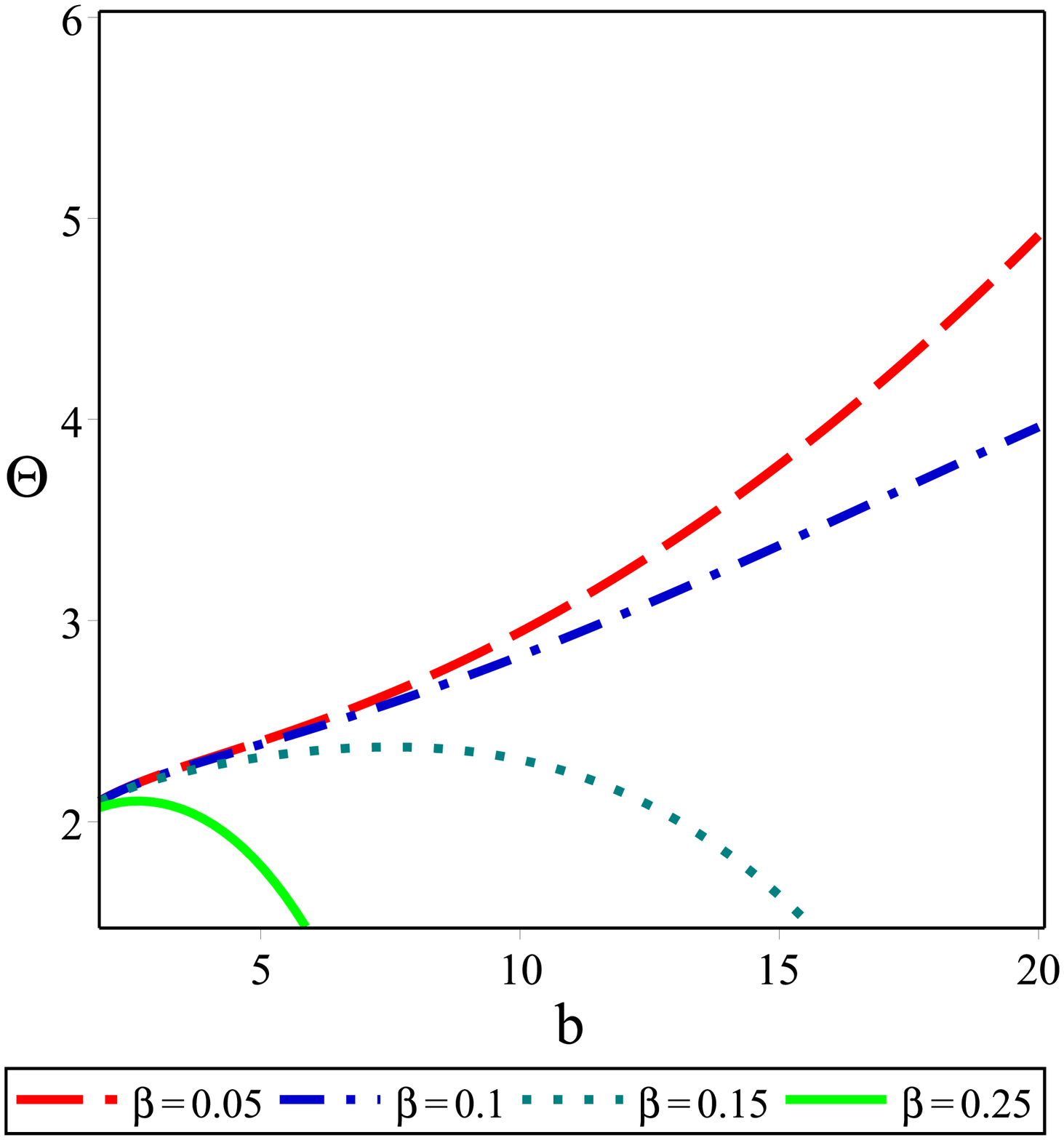}}
 \subfigure[~$\alpha=\beta=Q=0.2$ ]{
    \label{Fig12g}    \includegraphics[width=0.22\textwidth]{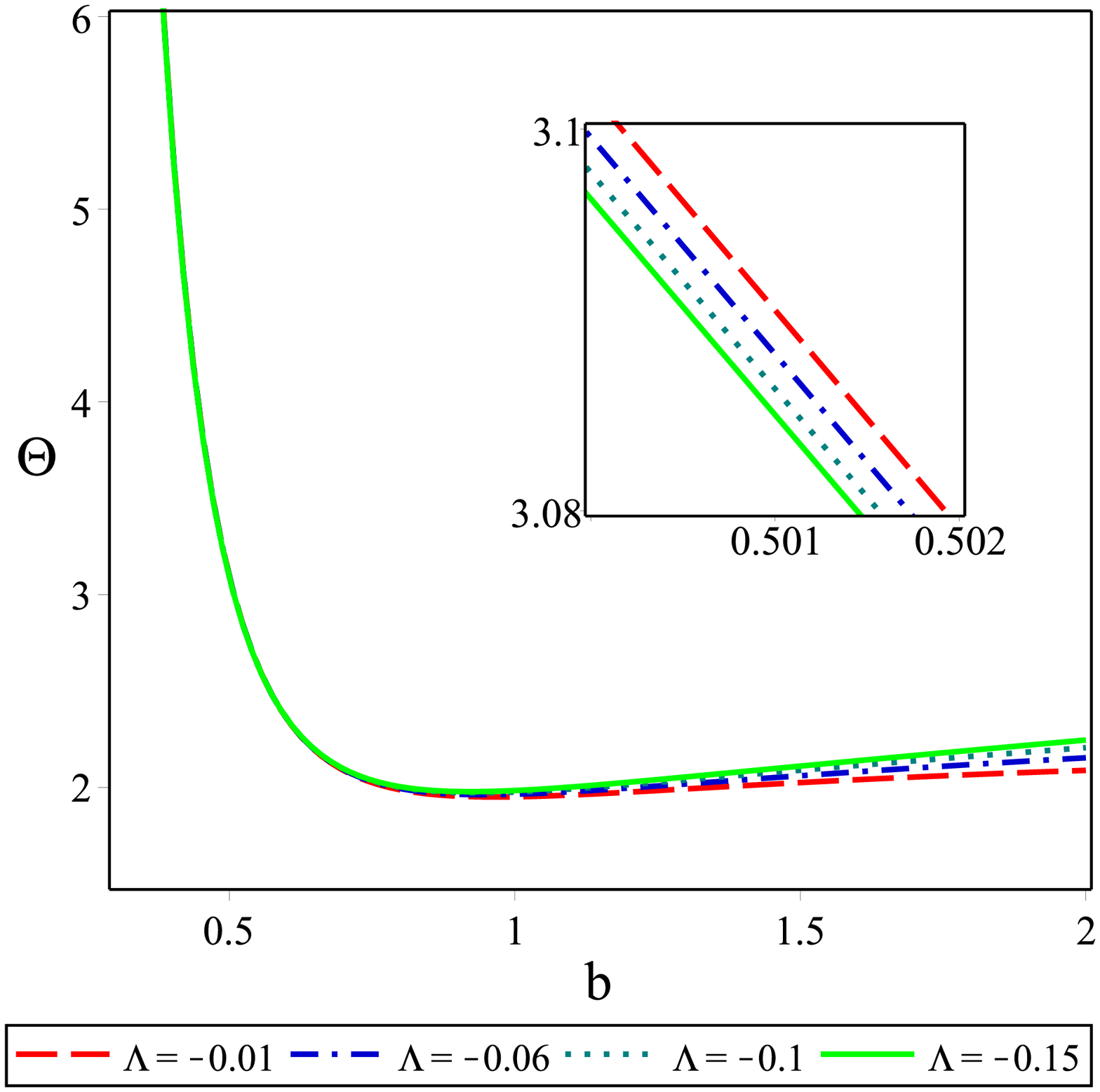}}
 \subfigure[~$\alpha=\beta=Q=0.2$ ]{
    \label{Fig12h}    \includegraphics[width=0.22\textwidth]{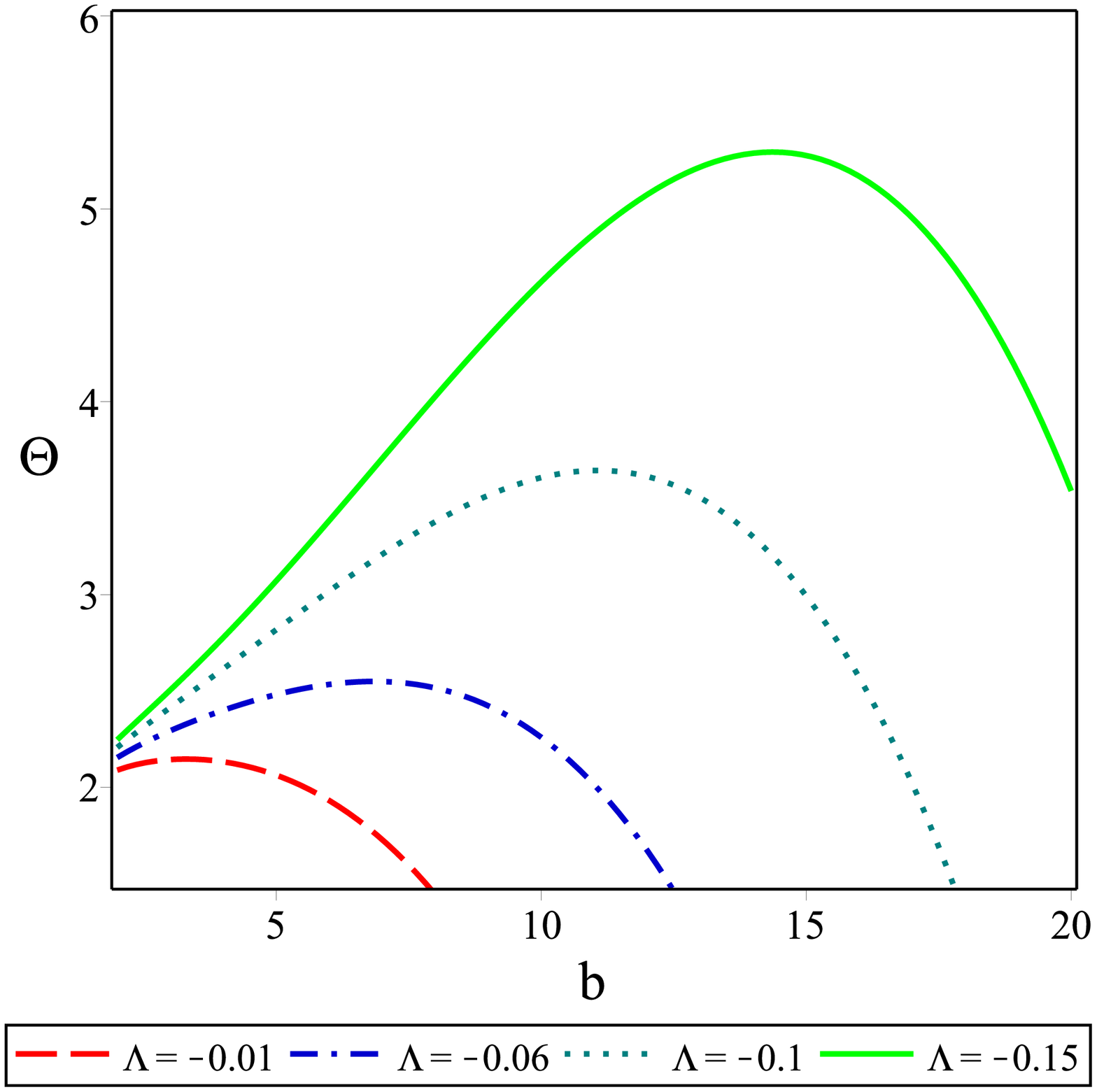}}
\caption{The behavior of $ \Theta $ with respect to the impact parameter $b$ for $ M=1 $ and different values of $ Q $, $ \beta $, $ \alpha $ and $\Lambda$. See the text for more details.}
\label{Fig12}
\end{figure*}

\section{Deflection angle using null geodesics}\label{sec:V}

In this section, we study the light deflection around the black hole solution (\ref{function-a}).  Employing the null geodesics method \cite{Chandrasekhar,Weinberg,Kocherlakota,WJaved}, one can determine the total deflection $ \Theta $ by the following relation
\begin{equation}
\Theta=2\int_{b}^{\infty}\Big\vert\frac{d\phi}{dr}\Big\vert dr -\pi
,  \label{EqDAn}
\end{equation}
where $b$ is the impact parameter, defined as $ b\equiv L/E $. Using Eq. (\ref{EqDAn}), the deflection angle is obtained as
%
\begin{eqnarray}
\Theta &=&\frac{M\alpha (24Mb+7Q^{2})}{42b^{5}}+\frac{(2\alpha\Lambda -3)(2Q^{2}+15Mb)}{90b^{2}}
	\nonumber  \\
&&-\frac{2b^{2}\beta^{2}(2\alpha \Lambda -3)}{27}{~}_{2}F_{1}\left( -\frac{3}{4},-\frac{1}{2},\frac{5}{4},-\frac{Q^{2}}{\beta^{2} b^{4}}\right) \nonumber
\end{eqnarray} 
\begin{eqnarray} 
&&-\frac{2\alpha b^{2}\beta^{4}(5Q^{2}+24Mb)}{189Q^{2}}{~}_{2}F_{1}\left( -\frac{7}{4},-\frac{3}{2},-\frac{3}{4},-\frac{Q^{2}}{\beta^{2} b^{4}}\right)
	\nonumber  \\
&&+\frac{7}{3} +\frac{\Lambda b^{2}(\alpha\Lambda -3)}{9}+\frac{2\beta^{2}}{27b^{2}}\big[(2\alpha Q^{2}+12\alpha Mb
	\nonumber  \\
&& \qquad +b^{4}(2\alpha \Lambda -3)\big]{~}_{2}F_{1}\left( -\frac{3}{4},-\frac{1}{2},\frac{1}{4},-\frac{Q^{2}}{\beta^{2} b^{4}}\right)
.\label{EqIDA}
\end{eqnarray}

Figure \ref{Fig12} depicts the impact of the black hole parameters on the deflection angle. More specifically, Figs. \ref{Fig12a} and \ref{Fig12b} illustrate the effect of electric charge on $ \Theta $. One verifies that $ \Theta $ is a decreasing (an increasing) function of $ Q $ for small (large) values of the impact parameter. Figures \ref{Fig12c} and \ref{Fig12d} indicate the behavior of $ \Theta $ by varying the GB parameter. As we see, for small values of $ b $, the effect of $ \alpha $ is to increase the deflection angle, whereas for large $ b $, its effect is to decrease $ \Theta $. This shows that the effect of $ Q $ and $ \alpha$ on the deflection angle is opposed to each other. Now, analysing Figs. \ref{Fig12e} and \ref{Fig12f}, we verify that for all values of the impact parameter, increasing the BI parameter leads to the decreasing of $ \Theta $. Regarding the cosmological constant effect, we see from Figs. \ref{Fig12g} and \ref{Fig12h}, that its effect is similar to that of the GB parameter. Using Fig. \ref{Fig12}, one can also analyze the behavior of $ \Theta $ by varying the impact parameter. We verify that for small and large values of $ b $, the deflection angle is a decreasing function of $ b $, whereas for intermediate values, it is an increasing function; except for low values of $\beta$, where the increasing behavior is observed for large values of $b$ as well (see dashed line in Fig. \ref{Fig12f}).

\section{Summary and Conclusion}\label{sec:conclusion}

Motivated by interesting properties of black hole solutions in the recently proposed regularized 4D EGB theory of gravity, we considered Born-Infeld black holes and studied their dynamic features including the shadow size, energy emission rate, quasinormal modes and deflection angle. We first investigated the shadow behavior and photon sphere and examined how the black hole parameters affect  them. The results showed that the GB parameter decrease both the photon sphere radius and shadow size, albeit, these quantities are not that sensitive to this parameter. Studying the effect of electric charge, we noticed that its effect was similar to that of GB coupling constant. However, the effect of the cosmological constant differs. Although it decreases  the photon sphere radius, it has an increasing effect on the shadow size. Then, we performed an investigation regarding the energy emission rate and analyzed the effective roles of these parameters on this optical quantity. We found that all these parameters have a decreasing contribution on the emission rate, namely, the emission of particles around the black hole decreases by increasing these parameters. This revealed the fact that such black holes have a longer lifetime if they are located in a background with higher curvature or in a stronger electric field.

Furthermore,  we employed the connection between the radius of the shadow and quasinormal modes and studied small perturbations around the black holes. We found that: i) As the coupling constants increase the  real part (absolute value of the imaginary part) of the quasinormal frequencies increases (decreases). This means that
QNMs oscillate faster and decay slower around the black holes with powerful coupling. ii) The effect of the cosmological constant is to decrease both real and imaginary parts of the QNMs. In other words, scalar perturbations have less energy for oscillation and decay slower in a higher curvature background. iii) The real (imaginary) part of QNM is an increasing (decreasing) function of the electric charge. In fact, the scalar field perturbations in the presence of electric charge oscillate faster and decay more slowly compared to neutral black holes.

Finally, we presented a study in the context of the gravitational lensing of light around these black holes. Depending on the choice of the BH parameters, we observed different behaviors. For small and large values of $ b $, the deflection angle is a decreasing function of $ b $, whereas for intermediate values, it is an increasing function. Relative to the impact of electric charge, we found that its effect is dependent on the values of the impact parameter, such that for small values of $b$, $\Theta$ is a decreasing function of $Q$, however, for large values, the effect of $ Q $ is opposite. We also found that the effects of the GB parameter and the cosmological constant on the deflection angle are opposed to that of the electric charge. Regarding the influence of BI parameter, it had a decreasing effect for all values of $ b $, although quite insignificant.

Here, we studied the optical features of the black hole solution in a modified gravity. There are other interesting solutions such as charged accelerating AdS black holes in $f(R)$ gravity \cite{Zhang:a}, black bounce solutions \cite{Lobo} and regular black holes with an asymptotically Minkowski core \cite{Berry:a} which we are currently studying. The results will appear elsewhere.

\begin{acknowledgments}
We thank the referee for constructive comments which helped us improve the paper significantly.
MKZ would like to thank Shahid Chamran University of Ahvaz, Iran for supporting this work.
FSNL acknowledges support from the Funda\c{c}\~{a}o para a Ci\^{e}ncia e a Tecnologia (FCT) Scientific Employment Stimulus contract with reference CEECIND/04057/2017. FSNL also thanks funding from the FCT research grants No. UID/FIS/04434/2020, No. PTDC/FIS-OUT/29048/2017 and No. CERN/FIS-PAR/0037/2019.
\end{acknowledgments}


\end{document}